\documentclass[12pt]{article}

\def\square{\kern1pt\vbox{\hrule height 1.2pt\hbox{\vrule width 1.2pt\hskip 3pt
   \vbox{\vskip 6pt}\hskip 3pt\vrule width 0.6pt}\hrule height 0.6pt}\kern1pt}

\begin{document}

\begin{titlepage}

\begin{flushright}
ITP-UU-11/18 \\ SPIN-11/13 \\ CCTP-2011-15 \\ UFIFT-QG-11-03
\end{flushright}

\begin{center}
{\bf The Graviton Propagator in de Donder Gauge on de Sitter Background}
\end{center}

\begin{center}
S. P. Miao$^*$
\end{center}

\begin{center}
\it{Institute for Theoretical Physics \& Spinoza Institute, Utrecht
University \\ Leuvenlaan 4, Postbus 80.195, 3508 TD Utrecht, NETHERLANDS}\\
\end{center}

\begin{center}
N. C. Tsamis$^{\dagger}$
\end{center}

\begin{center}
\it{Institute of Theoretical Physics \& Computational Physics, Dept. of Physics \\
University of Crete, GR-710 03 Heraklion, HELLAS}
\end{center}

\begin{center}
R. P. Woodard$^{\ddagger}$
\end{center}

\begin{center}
\it{Department of Physics, University of Florida \\
Gainesville, FL 32611, UNITED STATES}
\end{center}

\begin{center}
ABSTRACT
\end{center}
We construct the graviton propagator on de Sitter background in
exact de Donder gauge. We prove that it must break de Sitter
invariance, just like the propagator of the massless, minimally
coupled scalar. Our explicit solutions for its two scalar structure
functions preserve spatial homogeneity and isotropy so that the
propagator can be used within the larger context of inflationary
cosmology, however, it is simple to alter the residual symmetry.
Because our gauge condition is de Sitter invariant (although no
solution for the propagator can be) renormalization should be
simpler using this propagator than one based on a noncovariant
gauge. It remains to be seen how other computational steps compare.

\begin{flushleft}
\end{flushleft}

\begin{flushleft}
$^*$ e-mail: S.Miao@uu.nl \\
$^{\dagger}$ e-mail: tsamis@physics.uoc.gr \\
$^{\ddagger}$ e-mail: woodard@phys.ufl.edu
\end{flushleft}

\end{titlepage}

\section{Introduction}

There are bound to be interesting lessons when two plausible
arguments lead to opposite conclusions. The conflict we have in mind
concerns the graviton propagator on de Sitter background and the
arguments about it derive, respectively, from cosmology and
mathematical physics:
\begin{itemize}
\item{From the perspective of cosmology, the graviton propagator
cannot be de Sitter invariant because the unique Fourier mode sum
for an invariant propagator is infrared divergent as a consequence
of the scale invariance of the tensor power spectrum \cite{MTW1}.}
\item{From the perspective of mathematical physics, the graviton
propagator must be de Sitter invariant because explicit, de Sitter
invariant solutions for the graviton propagator equation have been
obtained when covariant gauge fixing terms are added to the action
\cite{INVPROP}.}
\end{itemize}

The cosmology argument turns out to be correct. Seeing this teaches:
\begin{itemize}
\item{There is a topological obstacle to adding covariant gauge
fixing terms on any manifold, and for any gauge theory, which
possesses a linearization instability \cite{MTW1}.}
\item{One can impose covariant gauges in which the field obeys
exact conditions, but previous solutions employed analytic
continuation techniques that incorrectly subtract off power law
infrared divergences \cite{MTW2}.}
\item{Solutions exist to the propagator equation which do not
correspond to propagators in the sense of being the expectation
value of the time-ordered product of two fields in the presence of
some state \cite{TW1}.\footnote{It has been conjectured that this
shows up in mathematical physics as a violation of reflection
positivity \cite{Marolf}.}}
\end{itemize}
The first point occurs even for flat space electromagnetism on the
manifold $T^3 \times R$: the invariant equations' linearization
instability requires the total charge to vanish, whereas the Feynman
gauge equations can be solved for any charge. The second point is
familiar to everyone who has encountered the automatic subtraction
of dimensional regularization. And a trivial example of the third
point comes from multiplying the entirely real, retarded propagator
by a factor of $i$.

These insights resolve a number of puzzles in the literature. For
example, employing the Feynman gauge fixing term for scalar quantum
electrodynamics on de Sitter \cite{AJ} produces a one loop
self-mass-squared which possesses on-shell singularities \cite{KW1}.
These singularities seem to be the quantum field theory
representation of what one would expect classically from an $A_0
J^0$ interaction energy in view of the erroneous temporal growth of
$A_0$ in Feynman gauge. The simplest noncovariant gauge \cite{RPW}
fails to show on-shell singularities \cite{KW1}. Nor is there any
problem using the de Sitter invariant, Lorentz gauge propagator
\cite{TW2,PTW1}. The conclusion for de Sitter electromagnetism is
therefore that one must avoid adding covariant gauge fixing terms,
but no physical breaking of de Sitter invariance occurs.

The situation for gravitons is different owing to infrared
divergences. It has long been noted that certain discrete choices of
the two covariant gauge fixing parameters result in infrared
divergences if one insists on a de Sitter invariant solution
\cite{IAEM,Folacci}. These choices had been dismissed as unphysical,
``singular gauges'' which must simply be avoided \cite{Higuchi}.
However, we can now see that they are precisely the cases for which
the order of the omnipresent infrared divergence in the formal, de
Sitter invariant mode sum changes from power law to logarithmic
\cite{MTW2}. The power law infrared divergences of other choices
were automatically --- but incorrectly --- subtracted by analytic
regularization techniques to produce solutions of the propagator
equations that are not true propagators. The correct procedure in
all cases is to allow free gravitons to resolve their infrared
problem by breaking de Sitter invariance.

The purpose of this note is to construct the graviton propagator in
an allowed covariant gauge, without employing analytic continuation
techniques to subtract off infrared divergences. Our procedure is to
express the propagator in terms of covariant projectors acting on
scalar structure functions, without making any assumption about the
eventual de Sitter invariance of the result. These structure
functions obey completely de Sitter invariant equations, but they
fail to possess de Sitter invariant solutions on account of infrared
divergences. The procedure is so general that we implement it as
well for a vector particle of general mass $M_V$, and check that it
agrees with the known de Sitter invariant solutions \cite{MTW2} for
$M_V^2 > -2(D-1) H^2$ in the transverse sector and $M_V^2 > 0$ in
the longitudinal sector. When de Sitter breaking must occur we have
chosen to give explicit solutions which preserve the symmetries of
homogeneity and isotropy that are relevant to cosmology. However,
our equations for the structure functions are invariant, so one can
easily derive solutions which respect any of the allowed subgroups.

Our notation is laid out in section 2. Section 3 presents a general
treatment for minimally coupled scalars of any mass $M_S$. In
section 4 we solve for the propagator of a vector with general mass
$M_V$, including longitudinal and transverse parts. Section 5
applies the same technique to solve for the graviton propagator in
de Donder gauge. Our results are summarized and discussed in section
6.

Because this work represents a long and mostly technical exercise we
have thought it right to briefly discuss the physical motivation.
The point is to facilitate the study of quantum effects from the
epoch of primordial inflation for which the de Sitter geometry
provides an excellent paradigm.\footnote{Just how good can be
quantified using the deceleration parameter $q(t) =
-a\ddot{a}/\dot{a}^2$, which measures minus the fractional cosmic
acceleration. Its value for de Sitter is $q = -1$, and the threshold
between inflation and deceleration occurs at $q=0$. If one assumes
single scalar inflation then the measured result for the scalar
amplitude \cite{WMAP}, and the bound on the tensor-to-scalar ratio
\cite{WMAP}, imply $95\%$ confidence that $q(t) < -0.986$ when the
largest observable perturbations experienced first horizon crossing
\cite{KOW}. Because this would have been near the end of inflation,
when $q(t)$ was growing, most of the inflationary epoch was likely
even closer to de Sitter.} The source of these effects is particle
production. The small amount of particle production which has long
been known to occur in an expanding universe \cite{old} becomes
explosive during inflation for any particle which is both massless
and not conformally invariant on the classical level \cite{Parker}.
This includes massless, minimally coupled scalars and gravitons
\cite{Grishchuk}. Of course this phenomenon is the origin of the
tensor \cite{Starobinsky} and scalar \cite{Mukhanov} perturbations
which are such an exciting tree order prediction of inflation. Our
motivation is getting at the fascinating loop effects which should
also be present.

There have been extensive studies of the quantum loop effects from
inflation producing massless, minimally coupled scalars. In three
models there are complete, dimensionally regulated and fully
renormalized results:
\begin{itemize}
\item{For a real scalar with a quartic self-interaction there are
one and two loop results for the expectation value of the stress
tensor \cite{OW} and the self-mass-squared \cite{BOW}. These show
that inflationary particle production pushes the scalar up its
potential, which increases the vacuum energy and leads to a
violation of the weak energy condition on cosmological scales
without any instability.}
\item{For a massless fermion which is Yukawa-coupled to a real
scalar there are one loop results for the fermion self-energy
\cite{PWGP}, for the scalar self-mass-squared \cite{DW}, and for the
effective potential \cite{MW1}. There is also a two loop computation
of the coincident vertex function \cite{MW1}. These results show
that the inflationary production of scalars endows super-horizon
fermions with a mass, which decreases the vacuum energy without
bound in such a way that the universe eventually undergoes a Big Rip
singularity \cite{MW1}.}
\item{For scalar quantum electrodynamics there are one loop results
for the vacuum polarization \cite{PTW} and the scalar
self-mass-squared \cite{KW1,PTW1}, and there are two loop results
for coincident scalars \cite{PTW1}, for coincident field strength
tensors \cite{PTW2}, and for the expectation value of the stress
tensor \cite{PTW2}. These results show that the inflationary
production of charged scalars causes super-horizon photons develop a
mass, while the scalar remains light and the vacuum energy decreases
slightly \cite{PTW3}.}
\end{itemize}

Scalar effects are simpler than those from gravitons because there
is no issue about general coordinate invariance. They are also
generally stronger because they can avoid derivative interactions.
However, scalar effects are correspondingly less universal and less
reliable because they depend upon the existence of light, minimally
coupled scalars at inflationary scales. In four models with
gravitons there are complete, dimensionally regulated and fully
renormalized results:
\begin{itemize}
\item{For pure quantum gravity the graviton 1-point function has
been computed at one loop order \cite{TW3}. This result shows that
the effect of inflationary gravitons at one loop order is a slight
increase in the cosmological constant.}
\item{For quantum gravity plus a massless fermion the fermion
self-energy has been computed at one loop order \cite{MW2}. This
result shows that spin-spin interactions with inflationary gravitons
drive the fermion field strength up by an amount that increases
without bound \cite{MW3}.}
\item{For quantum gravity plus a massless, minimally coupled scalar
there are one loop computations of the scalar self-mass-squared
\cite{KW2} and the graviton self-energy \cite{PW}. The scalar
effective field equations reveal that the scalar kinetic energy
redshifts too rapidly for there to be a significant interaction with
inflationary gravitons \cite{KW2}. The effects of inflationary
scalars on dynamical gravitons, and on the force of gravity, are
still under study \cite{PW}.}
\item{The nonlinear sigma model has been exploited to better
understand the derivative interactions of quantum gravity
\cite{TW4}, and explicit two loop results have been obtained for the
expectation value of the stress tensor \cite{KK}.}
\end{itemize}
There are also a variety of other, sometimes less complete results,
including:
\begin{itemize}
\item{In pure quantum gravity, a very early, approximate computation
of the one loop graviton 1-point function was made \cite{Ford}, as
well as a later evaluation using adiabatic regularization
\cite{Finelli}. Momentum cutoff computations of the one loop
graviton self-energy \cite{TW5} and the two loop graviton 1-point
function \cite{TW6} have also been done.}
\item{For gravity plus a scalar the one loop scalar contribution to the
noncoincident (and hence unregulated) graviton self-energy has been
computed \cite{Albert}.}
\item{In scalar-driven inflation there have been computations of the
one loop back-reaction \cite{back}, culminating in the realization
that a physical measure of the expansion rate shows no significant
effect at one loop order \cite{measure}. (This issue is still open
at two loop order, and as well in pure gravity \cite{puremeasure}.)
There has also been a vast amount of work on loop corrections to the
scalar power spectrum \cite{many}, including corresponding work on
how to correctly quantify effects \cite{gauge}, and a powerful
theorem by Weinberg which limits the maximum rate at which
corrections can display secular growth \cite{Weinberg}. Recently
there has been renewed attention to the problem of untangling
infrared effects from ultraviolet divergences \cite{UVIR}.}
\item{In gravity plus generic matter much interest has been
devoted to the recent proposal by Polyakov \cite{Polyakov}
(following numerous antecedents \cite{oldclaims,Ford,IAEM,TW6,back})
that runaway particle production may destabilize de Sitter space
\cite{destab}.}
\end{itemize}

\section{Notation}

In the coming sections we shall study the de Sitter background
propagators of three kinds of fields: minimally coupled scalars with
arbitrary mass $M_S$, vectors with arbitrary mass $M_V$, and
gravitons. The respective Lagrangians are,
\begin{eqnarray}
\mathcal{L}_{S} & = & -\frac12 \partial_{\mu} \varphi \partial_{\nu}
\varphi g^{\mu\nu} \sqrt{-g} - \frac12 M_S^2 \varphi^2 \sqrt{-g} \;
, \label{scalarL} \\
\mathcal{L}_{V} & = & -\frac12 \partial_{\mu} A_{\rho}
\partial_{\nu} A_{\sigma} g^{\mu\nu} g^{\rho\sigma} \sqrt{-g}
-\frac12 \Bigl[(D \!-\! 1) H^2 \!+\! M_V^2 \Bigr] A_{\rho}
A_{\sigma} g^{\rho\sigma} \sqrt{-g} \; , \label{vectorL} \\
\mathcal{L}_G & = & \frac1{16 \pi G} \Bigl[R - (D\!-\!2) (D\!-\!1)
H^2\Bigr] \sqrt{-g} \; . \label{gravitonL}
\end{eqnarray}
Here $D$ is the dimension of spacetime, $H$ is the Hubble constant
(which gives the cosmological constant $(D \!-\! 1) H^2 = \Lambda$)
and $G$ is Newton's constant. We make no assumption that the vector
is transverse, although the form of its mass term in (\ref{vectorL})
obviously derives from partially integrating and commuting covariant
derivatives in the Maxwell Lagrangian, and then adding a spurious
longitudinal kinetic term. The propagator of such a field appears in
projection operators, even though the associated field cannot be
dynamical.

We define the graviton field as the perturbation of the full metric
$g_{\mu\nu}(x)$ about its background value $\overline{g}_{\mu\nu}$,
\begin{equation}
h_{\mu\nu}(x) \equiv g_{\mu\nu}(x) - \overline{g}_{\mu\nu}(x) \; .
\label{graviton}
\end{equation}
Once this definition has been made, there is no more point to
distinguishing the background from the full metric so we drop the
overbar and refer to the de Sitter background as simply
$g_{\mu\nu}(x)$. Graviton indices are raised and lowered using this
background field, for example, $h^{\mu}_{~\nu} \equiv g^{\mu\rho}
h_{\rho\nu}$. Covariant derivative operators $D_{\mu}$ and other
geometrical quantities are similarly constructed with respect to the
background. Of special importance is the Lichnerowicz operator
which, when simplified using the de Sitter result for the curvature
$R_{\rho\sigma\mu\nu} = H^2 (g_{\mu \rho} g_{\nu\sigma} -
g_{\mu\sigma} g_{\nu\rho})$, takes the form,
\begin{eqnarray}
\lefteqn{\mathbf{D}^{\mu\nu\rho\sigma} \equiv D^{(\rho} g^{\sigma)
(\mu} D^{\nu)} -\frac12 \Bigl[ g^{\mu\nu} D^{\rho} D^{\sigma} \!+\!
g^{\rho\sigma} D^{\mu} D^{\nu} \Bigr] } \nonumber \\
& & \hspace{2cm} + \frac12 \Bigl[ g^{\mu\nu} g^{\rho\sigma} \!-\!
g^{\mu (\rho} g^{\sigma) \nu}\Bigr] \square + (D\!-\!1) \Bigl[
\frac12 g^{\mu\nu} g^{\rho\sigma} \!-\! g^{\mu (\rho} g^{\sigma)
\nu} \Bigr] H^2 \; . \qquad \label{Lichnerowicz}
\end{eqnarray}
Here and henceforth parenthesized indices are symmetrized, and
$\square \equiv g^{\mu\nu} D_{\mu} D_{\nu}$ is the covariant
d'Alembertian operator. With the help of (\ref{Lichnerowicz}) we can
express the free part of the gravitational Lagrangian
(\ref{gravitonL}) in a convenient form,
\begin{equation}
\mathcal{L}_{G} = \frac{(D \!-\! 1) H^2}{8 \pi G} \sqrt{-g} + \Bigl(
{\rm Surface\ Term}\Bigr) - \frac12 h_{\mu\nu}
\mathbf{D}^{\mu\nu\rho\sigma} h_{\rho\sigma} \sqrt{-g} + O(h^3) \; .
\end{equation}

Much of our work will be valid in any coordinate realization of de
Sitter space. However, when breaking de Sitter is necessary we shall
always do so on the $D$-dimensional open conformal submanifold in
which de Sitter can be imagined as a special case of the larger
class of homogeneous, isotropic and spatially flat geometries
relevant to cosmology. A spacetime point $x^{\mu} = (x^0, x^i)$
takes values in the ranges,
\begin{equation}
-\infty < x^0 < 0 \quad {\rm and} \quad  -\infty < x^i < +\infty
\quad {\rm for} \quad i = 1,\ldots,(D\!-\!1) \; .
\end{equation}
In these coordinates the invariant element is,
\begin{equation}
ds^2 \equiv g_{\mu\nu} dx^{\mu} dx^{\nu} = a_x^2 \Bigl[-(dx^0)^2 +
d\vec{x} \!\cdot\! d\vec{x} \Bigr] = a_x^2 \eta_{\mu\nu} dx^{\mu}
dx^{\nu}\; ,
\end{equation}
where $\eta_{\mu\nu}$ is the Lorentz metric and $a_x \equiv -1/Hx^0$
is the scale factor.

Although important de Sitter breaking occurs, it turns out that the
vast majority of our propagator is de Sitter invariant. This
suggests to express it in terms of the de Sitter invariant length
function $y(x;z)$,
\begin{equation}
y(x;z) \equiv a_x a_z H^2 \Biggl[ \Bigl\Vert \vec{x} \!-\! \vec{z}
\Bigr\Vert^2 - \Bigl(\vert x^0 \!-\! z^0\vert \!-\! i
\epsilon\Bigr)^2 \Biggr]\; . \label{ydef}
\end{equation}
Except for the factor of $i\epsilon$ (whose purpose is to enforce
Feynman boundary conditions) the function $y(x;z)$ is closely
related to the invariant length $\ell(x;z)$ from $x^{\mu}$ to
$z^{\mu}$,
\begin{equation}
y(x;z) = 4 \sin^2\Bigl( \frac12 H \ell(x;z)\Bigr) \; .
\end{equation}

With this de Sitter invariant quantity $y(x;z)$, we can form a
convenient basis of de Sitter invariant bi-tensors. Note that
because $y(x;z)$ is de Sitter invariant, so too are covariant
derivatives of it. With the metrics $g_{\mu\nu}(x)$ and
$g_{\mu\nu}(z)$, the first three derivatives of $y(x;z)$ furnish a
convenient basis of de Sitter invariant bi-tensors \cite{KW1},
\begin{eqnarray}
\frac{\partial y(x;z)}{\partial x^{\mu}} & = & H a_x \Bigl(y
\delta^0_{\mu}
\!+\! 2 a_z H \Delta x_{\mu} \Bigr) \; , \label{dydx} \\
\frac{\partial y(x;z)}{\partial z^{\nu}} & = & H a_z \Bigl(y
\delta^0_{\nu}
\!-\! 2 a_x H \Delta x_{\nu} \Bigr) \; , \label{dydz} \\
\frac{\partial^2 y(x;z)}{\partial x^{\mu} \partial z^{\nu}} & = &
H^2 a_x a_z \Bigl(y \delta^0_{\mu} \delta^0_{\nu} \!+\! 2 a_z H
\Delta x_{\mu} \delta^0_{\nu} \!-\! 2 a_x \delta^0_{\mu} H \Delta
x_{\nu} \!-\! 2 \eta_{\mu\nu}\Bigr) \; . \qquad \label{dydxdx'}
\end{eqnarray}
Here and subsequently we define $\Delta x_{\mu} \equiv \eta_{\mu\nu}
(x \!-\!z)^{\nu}$. Acting covariant derivatives generates more basis
tensors, for example \cite{KW1},
\begin{eqnarray}
\frac{D^2 y(x;z)}{Dx^{\mu} Dx^{\nu}}
& = & H^2 (2 \!-\!y) g_{\mu\nu}(x) \; , \label{covdiv} \\
\frac{D^2 y(x;z)}{Dz^{\mu} Dz^{\nu}} & = & H^2 (2 \!-\!y)
g_{\mu\nu}(z) \; .
\end{eqnarray}
The contraction of any pair of the basis tensors also produces more
basis tensors \cite{KW1},
\begin{eqnarray}
g^{\mu\nu}(x) \frac{\partial y}{\partial x^{\mu}} \frac{\partial
y}{\partial x^ {\nu}} & = & H^2 \Bigl(4 y - y^2\Bigr) =
g^{\mu\nu}(z) \frac{\partial y}{
\partial z^{\mu}} \frac{\partial y}{\partial z^{\nu}} \; ,
\label{contraction1}\\
g^{\mu\nu}(x) \frac{\partial y}{\partial x^{\nu}} \frac{\partial^2
y}{
\partial x^{\mu} \partial z^{\sigma}} & = & H^2 (2-y) \frac{\partial y}{
\partial z^{\sigma}} \; ,
\label{contraction2}\\
g^{\rho\sigma}(z) \frac{\partial y}{\partial z^{\sigma}}
\frac{\partial^2 y}{\partial x^{\mu} \partial z^{\rho}} & = & H^2
(2-y) \frac{\partial y}{\partial x^{\mu}} \; ,
\label{contraction3}\\
g^{\mu\nu}(x) \frac{\partial^2 y}{\partial x^{\mu} \partial
z^{\rho}} \frac{\partial^2 y}{\partial x^{\nu} \partial z^{\sigma}}
& = & 4 H^4 g_{\rho\sigma}(z) - H^2 \frac{\partial y}{\partial
z^{\rho}} \frac{\partial y}{\partial z^{\sigma}} \; ,
\label{contraction4}\\
g^{\rho\sigma}(z) \frac{\partial^2 y}{\partial x^{\mu}\partial
z^{\rho}} \frac{\partial^2 y}{\partial x^{\nu} \partial z^{\sigma}}
& = & 4 H^4 g_{\mu\nu}(x) - H^2 \frac{\partial y}{\partial x^{\mu}}
\frac{\partial y}{\partial x^{\nu}} \; . \label{contraction5}
\end{eqnarray}

\section{Scalar Propagators}

Scalar propagator equations play an important role in our analysis
because our strategy is to enforce the de Donder gauge condition,
without making assumptions about de Sitter invariance, using
covariant derivative projectors acting on scalar structure
functions. The graviton propagator equation will then be used to
infer invariant equations for these scalar structure functions. The
point of this section is to review and systematize previous work
\cite{MTW1,MTW2} about how to solve such equations. We begin giving
a general scalar propagator equation and explaining why infrared
divergences for $M_S^2 \leq 0$ preclude a de Sitter invariant
solution. We review the two fixes in the literature, and then give a
simple approximate implementation for our favorite one. The section
closes with some powerful results for integrating propagators.

One can see from (\ref{scalarL}) that the propagator of a minimally
coupled scalar with mass $M_S$ obeys the equation,
\begin{equation}
\Bigl[ \square - M_S^2\Bigr] i\Delta(x;z) = \frac{i \delta^D(x \!-\!
z)}{\sqrt{-g}} \; . \label{nuprop}
\end{equation}
The plane wave mode function corresponding to Bunch-Davies vacuum is
\cite{BD},
\begin{equation}
u_{\nu}(x^0,k) \equiv \sqrt{\frac{\pi}{4 H}} \; a_x^{-\frac{D-1}2}
\, H^{(1)}_{\nu}(-k x^0) \quad {\rm where} \quad \nu =
\sqrt{\Bigl(\frac{D \!-\!1}2\Bigr)^2 \!-\! \frac{M_S^2}{H^2}} \; .
\label{unu}
\end{equation}
The Fourier mode sum for the propagator on infinite space is,
\begin{eqnarray}
\lefteqn{i\Delta^{\rm dS}_{\nu}(x;z) = \int \!\!
\frac{d^{D-1}k}{(2\pi)^{D-1}} \, e^{i \vec{k} \cdot (\vec{x} -
\vec{z})} \Biggl\{ \theta(x^0 \!-\! z^0)
u_{\nu}(x^0,k) u^*_{\nu}(z^0,k) } \nonumber \\
& & \hspace{6cm} + \theta(z^0 \!-\! x^0) u_{\nu}(x^0,k)
u_{\nu}(z^0,k) \Biggr\} . \qquad \label{modesum}
\end{eqnarray}
The result is de Sitter invariant when the integral converges
\cite{CT,JMPW1},
\begin{eqnarray}
\lefteqn{i\Delta^{\rm dS}_{\nu}(x;z) } \nonumber \\
& & \hspace{-.5cm} = \frac{H^{D-2}}{(4\pi)^{\frac{D}2}}
\frac{\Gamma(\frac{D-1}2 \!+\! \nu) \Gamma(\frac{D-1}2 \!-\! \nu)}{
\Gamma(\frac{D}2)} \, \mbox{}_2 F_1\Bigl( \frac{D-1}2 \!+\! \nu,
\frac{D-1}2 \!-\! \nu; \frac{D}2;1 \!-\! \frac{y}4\Bigr) \; , \qquad \\
& & \hspace{-.5cm} = \frac{H^{D-2} \Gamma(\frac{D}2 \!-\! 1)}{(4
\pi)^{\frac{D}2}} \Biggl\{ \Bigl( \frac{4}{y}\Bigr)^{\frac{D}2 -1}
\mbox{}_2 F_1\Bigl(\frac12 \!+\!\nu,\frac12 \!-\! \nu ; 2 \!-\!
\frac{D}2 ; \frac{y}4\Bigr) \qquad \nonumber \\
& & + \frac{\Gamma(\frac{D-1}2 \!+\!\nu) \Gamma(\frac{D-1}2 \!-\!
\nu) \Gamma(1 \!-\! \frac{D}2)}{\Gamma(\frac12 \!+\! \nu)
\Gamma(\frac12 \!-\! \nu) \Gamma(\frac{D}2 \!-\! 1)} \, \mbox{}_2
F_1\Bigl(\frac{D-1}2 \!+\! \nu , \frac{D-1}2 \!-\! \nu ;
\frac{D}2 ; \frac{y}4\Bigr) \Biggr\} , \qquad \\
& & \hspace{-.5cm} = \frac{H^{D-2}}{(4 \pi)^{\frac{D}2}} \Biggl\{
\Gamma\Bigl(\frac{D}2 \!-\! 1\Bigr)
\Bigl(\frac{4}{y}\Bigr)^{\frac{D}2-1}
\nonumber \\
& & \hspace{2cm} - \frac{\Gamma(\frac{D}2) \Gamma(1 \!-\!
\frac{D}2)}{ \Gamma(\frac12 \!+\! \nu) \Gamma(\frac12 \!-\! \nu)}
\sum_{n=0}^{\infty} \Biggl[ \frac{\Gamma(\frac32 \!+\! \nu \!+\! n)
\Gamma(\frac32 \!-\! \nu \!+\! n)}{ \Gamma(3 \!-\! \frac{D}2 \!+\!
n) \, (n \!+\! 1)!} \Bigl(\frac{y}4
\Bigr)^{n - \frac{D}2 +2} \nonumber \\
& & \hspace{5cm} - \frac{\Gamma(\frac{D-1}2 \!+\! \nu \!+\! n)
\Gamma(\frac{D-1}2 \!-\! \nu \!+\! n)}{\Gamma(\frac{D}2 \!+\! n) \,
n!} \Bigl(\frac{y}4\Bigr)^n \Biggr] \Biggr\} . \qquad
\label{expansion}
\end{eqnarray}

The gamma function $\Gamma(\frac{D-1}2 -\nu + n)$ on the final line
of (\ref{expansion}) diverges for,
\begin{equation}
\nu = \Bigl(\frac{D\!-\!1}{2}\Bigr) + N \qquad \Longleftrightarrow
\qquad M_S^2 = - N(D \!-\!1 \!+\! N) H^2 \; . \label{logprob}
\end{equation}
Its origin can be understood by performing the angular integration
in the naive mode sum (\ref{modesum}) and then changing to the
dimensionless variable $\tau \equiv k/H\sqrt{a_x a_z}$,
\begin{eqnarray}
\lefteqn{ i\Delta^{\rm dS}_{\nu}(x;z) = \frac{ (a_x
a_z)^{-(\frac{D-1}2)}}{ 2^D \pi^{\frac{D-3}2} H} \int_0^{\infty}
\!\! dk \, k^{D-2} \Bigl( \frac12 k \Delta x\Bigr)^{-(\frac{D-3}2)}
J_{\frac{D-3}2}(k \Delta x) } \nonumber \\
& & \times \Biggl\{ \theta(x^0 \!-\! z^0) H_{\nu}^{(1)}(-k x^0)
H_{\nu}^{(1)}(-k z^0)^* + \theta(z^0 \!-\! x^0) \Bigl({\rm
conjugate}
\Bigr) \Biggr\} , \qquad \\
& & \hspace{-.5cm} = \frac{H^{D-2}}{2^D \pi^{\frac{D-3}2}}
\int_0^{\infty} \!\! d\tau \, \tau^{D-2} \Bigl( \frac12 \sqrt{a_x
a_z} \, H \Delta x \tau \Bigr)^{-(\frac{D-3}2)}
J_{\frac{D-3}2}\Bigl( \sqrt{a_x a_z} \,
H \Delta x \tau\Bigr) \nonumber \\
& & \times \Biggl\{ \theta(x^0 \!-\! z^0) H_{\nu}^{(1)}\Bigl(
\sqrt{\frac{a_z}{a_x}} \, \tau\Bigr)
H_{\nu}^{(1)}\Bigl(\sqrt{\frac{a_x}{a_z}} \, \tau\Bigr)^* +
\theta(z^0 \!-\! x^0) \Bigl({\rm conjugate}\Bigr) \Biggr\} . \qquad
\label{integral}
\end{eqnarray}
In these and subsequent expressions we define $\Delta x \equiv \Vert
\vec{x} \!-\! \vec{z}\Vert$. That the divergence at (\ref{logprob})
is infrared can be seen from the small argument expansion of the
Bessel function and from its relation to the Hankel function,
\begin{eqnarray}
J_{\nu}(x) & = & \sum_{n=0}^{\infty} \frac{ (-1)^n (\frac12 x)^{\nu
+2n}}{
n! \Gamma(\nu \!+\! n \!+\! 1)} \; , \\
H_{\nu}^{(1)}(x) & = & \frac{i \Gamma(\nu) \Gamma(1 \!-\! \nu)}{\pi}
\Bigl\{ e^{-i\nu \pi} J_{\nu}(x) \!-\! J_{-\nu}(x)\Bigr\} \; .
\end{eqnarray}
The small $\tau$ behavior of the integrand (\ref{integral}) derives
from three factors, the first being $\tau^{D-2}$. The second factor
takes the form,
\begin{equation}
\Bigl( \frac12 \sqrt{a_x a_z} \, H \Delta x
\tau\Bigr)^{-(\frac{D-3}2)} J_{\frac{D-3}2}\Bigl( \sqrt{a_x a_z} \,
H \Delta x \tau\Bigr) = \frac1{\Gamma(\frac{D-1}2)}
\sum_{n=0}^{\infty} C_1(n) \tau^{2n} \; .
\end{equation}
And the final factor from the Hankel functions is,
\begin{equation}
H_{\nu}^{(1)}\Bigl(\sqrt{\frac{a_z}{a_x}} \, \tau\Bigr)
H_{\nu}^{(1)}\Bigl(\sqrt{\frac{a_x}{a_z}} \, \tau\Bigr)^* = \frac{2
\Gamma(\nu) \Gamma(2 \nu)}{\pi^{\frac32} \Gamma(\nu \!+\! \frac12)
\tau^{2\nu} } \sum_{n=0}^{\infty} C_2(n) \tau^{2n} \; .
\end{equation}
Hence the small $\tau$ expansion of the integrand has the form,
\begin{eqnarray}
\lefteqn{\tau^{D-2} \times \frac1{\Gamma(\frac{D-1}2)}
\sum_{k=0}^{\infty} C_1(k) \tau^{2k} \times \frac{\Gamma^2(\nu)
2^{2\nu}}{\pi^2 \tau^{2\nu} }
\sum_{\ell=0}^{\infty} C_2(\ell) \tau^{2\ell} } \nonumber \\
& & \hspace{4cm} = \frac{2 \Gamma(\nu) \Gamma(2\nu)}{\pi^{\frac32}
\Gamma(\frac{D-1}2) \Gamma(\nu \!+\! \frac12)} \, \tau^{D-2-2\nu}
\sum_{n=0}^{\infty} C_3(n) \tau^{2n} \; . \qquad \label{IR}
\end{eqnarray}
The naive mode sum (\ref{modesum}) is infrared divergent for
\begin{equation}
D - 2 - 2\nu \leq -1 \qquad \Longleftrightarrow \qquad M_S^2 \leq 0
\; . \label{truediv}
\end{equation}
However, there will only be a {\it logarithmic} infrared divergence,
either from the leading term in (\ref{IR}) or from one of the series
corrections at $n = N$, if one has,
\begin{equation}
D - 2 - 2\nu + 2N = -1 \qquad \Longleftrightarrow \qquad M_S^2 = -N
(D \!-\! 1 \!+\! N) H^2 \; .
\end{equation}
This is precisely the condition (\ref{logprob}) for the formal, de
Sitter invariant mode sum (\ref{expansion}) to diverge.

The infrared divergence we have just seen was first noted in 1977
for the special case of $M_S = 0$ by Ford and Parker \cite{FP}. The
appearance of an infrared divergence signals that something is
unphysical about the quantity being computed. The correct response
to an infrared divergence is not to subtract it off, either
explicitly or implicitly with the automatic subtraction of some
analytic regularization technique. One must instead understand the
physical problem which caused the divergence and then fix that
problem. As we will see, the fix involves breaking de Sitter
invariance, which was realized in 1982 for the special case of $M_S
= 0$ \cite{classic}. Allen and Folacci later gave a rigorous proof
that de Sitter invariance must be broken \cite{AF}.

The divergence (\ref{truediv}) occurs because of the way the
Bunch-Davies mode functions (\ref{unu}) depend upon $k$ for small
$k$. The unphysical thing about having Bunch-Davies vacuum for
arbitrarily small $k$ is that no experimentalist can causally
enforce it (or any other condition) for super-horizon modes. This
has led to two fixes:
\begin{enumerate}
\item{One can continue to work on the spatial manifold $R^{D-1}$
but assume the initial state is released with its super-horizon
modes in some less singular condition \cite{AV}; or}
\item{One can work on the compact spatial manifold $T^{D-1}$
with its coordinate radius chosen so the initial state has no
super-horizon modes \cite{TW7}.}
\end{enumerate}
We will adopt the latter fix. This makes the mode sum discrete, but
the integral approximation should be excellent, and gives a simple
expression for the propagator which differs from (\ref{modesum})
only by an infrared cutoff at $k = H$.

From the preceding discussion we see that the infrared corrected
propagator $i\Delta(x;z)$  is just (\ref{integral}) with the lower
limit cutoff at $\tau = 1/\sqrt{a_x a_z}$,
\begin{eqnarray}
\lefteqn{i\Delta(x;z) = \frac{H^{D-2}}{2^D \pi^{\frac{D-3}2}}
\int_{\frac1{\sqrt{a_x a_z}}}^{\infty} \!\!\!\!\! d\tau \,
\tau^{D-2} \frac{J_{\frac{D-3}2}( \sqrt{a_x a_z} \, H \Delta x
\tau)}{(\frac12
\sqrt{a_x a_z} \, H \Delta x \tau)^{\frac{D-3}2} } } \nonumber \\
& & \times \Biggl\{ \theta(x^0 \!-\! z^0) H_{\nu}^{(1)}\Bigl(
\sqrt{\frac{a_z}{a_x}} \, \tau\Bigr)
H_{\nu}^{(1)}\Bigl(\sqrt{\frac{a_x}{a_z}} \, \tau\Bigr)^* +
\theta(z^0 \!-\! x^0) \Bigl({\rm conjugate}\Bigr) \Biggr\} . \qquad
\end{eqnarray}
Of course we can express the truncated integral as the full one
minus an integral over just the infrared,
\begin{equation}
\int_{\frac1{\sqrt{a_x a_z}}}^{\infty} \!\!\!\!\! d\tau =
\int_0^{\infty} \!\! d\tau - \int_0^{\frac1{\sqrt{a_x a_z}}} \!\!\!
d\tau \quad \Longleftrightarrow \quad i\Delta(x;z) \equiv
i\Delta^{\rm dS}_{\nu}(x;z) + \Delta^{\rm IR}_{\nu}(x;z) \; .
\end{equation}
In this case it does not matter if dimensional regularization is
used to evaluate both $i\Delta^{\rm dS}_{\nu}(x;z)$ and $\Delta^{\rm
IR}_{\nu}(x;z)$ because the errors we make at the lower limits will
cancel.

A further simplification is that $\Delta^{\rm IR}_{\nu}(x;z)$ only
needs to include the infrared singular terms which grow as $a_x a_z$
increases. These terms come entirely from the $J_{-\nu}$ parts of
the Hankel function and they are entirely real,
\begin{eqnarray}
\lefteqn{\Delta^{\rm IR}_{\nu}(x;z) =
-\frac{H^{D-2}}{(4\pi)^{\frac{D}2}} \frac{2 \Gamma(\nu)
\Gamma(2\nu)}{\Gamma(\nu \!+\! \frac12)} \int_0^{\frac1{\sqrt{a_x
a_z}}} \!\!\! d\tau \, \tau^{D-2} \frac{J_{\frac{D-3}2}( \sqrt{a_x
a_z} \, H \Delta x \tau)}{(\frac12
\sqrt{a_x a_z} \, H \Delta x \tau)^{\frac{D-3}2} } } \nonumber \\
& & \hspace{5.5cm} \times \frac{\Gamma^2(1\!-\!\nu)}{2^{2\nu}} \,
J_{-\nu}\Bigl( \sqrt{\frac{a_z}{a_x}} \, \tau\Bigr) J_{-\nu}\Bigl(
\sqrt{\frac{a_x}{a_z}} \, \tau\Bigr) \; . \qquad \label{homo}
\end{eqnarray}
The final result is \cite{JMPW2,MTW2},
\begin{eqnarray}
\lefteqn{\Delta^{\rm IR}_{\nu}(x;z) =
\frac{H^{D-2}}{(4\pi)^{\frac{D}2}} \frac{\Gamma(\nu)
\Gamma(2\nu)}{\Gamma(\frac{D-1}2)
\Gamma(\nu \!+\! \frac12)} } \nonumber \\
& & \hspace{.7cm} \times \sum_{N=0}^{\infty} \frac{(a_x a_z)^{\nu -
(\frac{D-1}2) - N}}{\nu \!-\! (\frac{D-1}2) \!-\! N} \sum_{n=0}^N
\Bigl( \frac{a_x}{a_z} \!+\! \frac{a_z}{a_x}\Bigr)^n \sum_{m=0}^{
[\frac{N-n}2]} C_{Nnm} (y \!-\!2)^{N-n-2m} \; , \qquad
\label{series}
\end{eqnarray}
where the coefficients $C_{Nnm}$ are,
\begin{eqnarray}
\lefteqn{C_{Nnm} = \frac{(-\frac14)^N}{m! n! (N \!-\!n \!-\! 2m)!}
\times \frac{\Gamma(\frac{D-1}2 \!+\! N \!+\! n \!-\!
\nu)}{\Gamma(\frac{D-1}2
\!+\! N \!-\! \nu)} } \nonumber \\
& & \hspace{2cm} \times
\frac{\Gamma(\frac{D-1}2)}{\Gamma(\frac{D-1}2 \!+\! N\!-\! 2m)}
\times \frac{\Gamma(1 \!-\!\nu)}{\Gamma(1 \!-\! \nu \!+\! n \!+\!
2m)} \times \frac{\Gamma(1 \!-\! \nu)}{\Gamma(1 \!-\! \nu \!+\! m)}
\; . \qquad \label{cdef}
\end{eqnarray}
Of course there is no point in extending the sum over $N$ to values
$N > \nu -(\frac{D-1}2)$ for which the exponent of $a_x a_z$ becomes
negative. Those terms rapidly approach zero, and they can be dropped
without affecting the propagator equation because they are
separately annihilated by $\square - M_s^2$.

It might be worried that the approximations made in deriving the
infrared correction do violence to delicate consistency relations in
quantum field theory, but this is not the case. For the $M_S = 0$
scalar renormalization has been successfully implemented at one and
two loop orders \cite{KW1,PTW1,OW,BOW,PWGP,MW1,PTW,PTW2,PTW3}.
Because the physical graviton polarizations obey the same mode
functions as massless, minimally coupled scalars \cite{Grishchuk},
one can also test the integral approximation with the graviton
propagator. There is no disruption of powerful consistency checks
such as the Ward identity at tree order \cite{TW8} and one loop
\cite{TW6}. Nor is there any problem with the allowed one loop
counterterms \cite{TW3,MW2,KW2,PW}.

It is worthwhile to summarize these results in the context of a
consistent notation. Consider a general scalar whose mass obeys
$M_s^2/H^2 = (\frac{D-1}2)^2 - b^2$. Its propagator $i\Delta_b(x;z)$
obeys the equation,
\begin{equation}
\Bigl[ \square_x + b^2 H^2 - \Bigl(\frac{D \!-\! 1}2\Bigr)^2 H^2
\Bigr] i \Delta_b(x;z) = \frac{i\delta^D( x\!-\! z)}{\sqrt{-g}} \; .
\label{bprop}
\end{equation}
We define the final result for $i\Delta_b(x;z)$ as the limit as
$\nu$ approaches $b$ of two functions which we wish to consider for
general index $\nu$. The first term in the sum is $i\Delta^{\rm
dS}_{\nu}(x;z)$ as defined by expression (\ref{expansion}). The
second term is $\Delta^{\rm IR}_{\nu}(x;z)$, as defined by
expression (\ref{series}), except that the sum over $N$ is cut off
at the largest nonnegative integer for which $N \leq b -
(\frac{D-1}2)$, with $\Delta^{\rm IR}_{\nu}(x;z)$ defined as zero
for $b < (\frac{D-1}2)$. Hence our final result is,
\begin{equation}
i\Delta_b(x;z) = \lim_{\nu \rightarrow b} \Bigl[ i\Delta^{\rm
dS}_{\nu}(x;z) + \Delta^{\rm IR}_{\nu}(x;z) \Bigr] \; .
\end{equation}

We shall make significant use of four special cases for which a
separate notation has been introduced:
\begin{eqnarray}
b_B = \Bigl(\frac{D \!-\! 3}2\Bigr) & \Longleftrightarrow & i\Delta_B(x;z) = B(y) \; , \\
b_A = \Bigl(\frac{D \!-\! 1}2\Bigr) & \Longleftrightarrow &
i\Delta_A(x;z) = A(y) + \delta A(a_x,a_z,y) \; , \label{Adef} \\
b_W = \Bigl(\frac{D \!+\! 1}2\Bigr) & \Longleftrightarrow &
i\Delta_W(x;z) = W(y) + \delta W(a_x,a_z,y) \; , \\
b_M = \frac12 \sqrt{(D \!-\! 1) (D \!+\! 7)} & \Longleftrightarrow &
i\Delta_M(x;z) = M(y) + \delta M(a_x,a_z,y) \; . \label{Mdef}
\end{eqnarray}
Although the $B$-type propagator is de Sitter invariant, its
$A$-type, $W$-type and $M$-type cousins have de Sitter breaking
parts,
\begin{eqnarray}
\delta A & = & k \ln(a_x a_z) \; , \label{dsbA} \\
\delta W & = & k \Biggl\{ (D\!-\!1)^2 a_x a_z -
\Bigl(\frac{D\!-\!1}2\Bigr) \ln(a_x a_z) (y \!-\! 2) -
\Bigl(\frac{a_x}{a_z} \!+\! \frac{a_z}{a_x}\Bigr) \Biggr\} \; ,
\qquad \\
\delta M & = & k_M \Biggl\{ \frac{ (a_x a_z)^{b_M - b_A}}{b_M \!-\!
b_A} - \frac{ (a_x a_z)^{b_M - b_A - 1}}{b_M \!-\! b_A \!-\! 1}
\times \frac{(y \!-\! 2)}{4 b_A} \nonumber \\
& & \hspace{5cm} + \frac{ (a_x a_z)^{b_M - b_A - 1}}{4 b_A (b_M
\!-\! 1)} \times \Bigl( \frac{a_x}{a_z} \!+\! \frac{a_z}{a_x}\Bigr)
\Biggr\} \; . \qquad
\end{eqnarray}
The constants $k$ and $k_M$ are,
\begin{equation}
k \equiv \frac{H^{D-2}}{(4\pi)^{\frac{D}2}} \, \frac{\Gamma(D \!-\!
1)}{ \Gamma(\frac{D}2)} \qquad , \qquad k_M \equiv
\frac{H^{D-2}}{(4\pi)^{\frac{D}2}} \frac{ \Gamma(b_M)
\Gamma(2b_M)}{\Gamma(b_A) \Gamma(b_M \!+\! \frac12)} \; .
\end{equation}
The main, de Sitter invariant parts of each propagator consist of a
few, potentially ultraviolet divergent terms (at $y=0$), plus an
infinite series. For the $M$-type propagator there are no
cancelations with the de Sitter breaking terms: just replace $\nu$
everywhere by $b_M$ in expression (\ref{expansion}) to find $M(y) =
i\Delta^{\rm dS}_{b_M}(x;z)$. However, there are cancelations when
this replacement is done for the $A$-type and $W$ propagators,
\begin{eqnarray}
\lefteqn{B(y) = \frac{H^{D-2}}{(4\pi)^{\frac{D}2}} \Biggl\{
\Gamma\Bigl(\frac{D}2 \!-\!1\Bigr)
\Bigl(\frac{4}{y}\Bigr)^{\frac{D}2 -1} }
\nonumber \\
& & \hspace{1.5cm} + \sum_{n=0}^{\infty} \Biggl[
\frac{\Gamma(n\!+\!\frac{D}2)}{(n \!+\! 1)!} \Bigl(\frac{y}4
\Bigr)^{n - \frac{D}2 +2} \!\!\!\!\! - \frac{\Gamma(n \!+\! D \!-\!
2)}{\Gamma(n \!+\! \frac{D}2)}
\Bigl(\frac{y}4 \Bigr)^n \Biggr] \Biggr\} , \qquad \label{DeltaB} \\
\lefteqn{A(y) = \frac{H^{D-2}}{(4\pi)^{\frac{D}2}} \Biggl\{
\Gamma\Bigl(\frac{D}2 \!-\!1\Bigr) \Bigl(\frac{4}{y}\Bigr)^{
\frac{D}2 -1} \!+\! \frac{\Gamma(\frac{D}2 \!+\! 1)}{\frac{D}2 \!-\!
2}
\Bigl(\frac{4}{y} \Bigr)^{\frac{D}2-2} \!+\! A_1 } \nonumber \\
& & \hspace{1.5cm} - \sum_{n=1}^{\infty} \Biggl[
\frac{\Gamma(n\!+\!\frac{D}2\!+\!1)}{(n\!-\!\frac{D}2\!+\!2) (n
\!+\! 1)!} \Bigl(\frac{y}4 \Bigr)^{n - \frac{D}2 +2} \!\!\!\!\! -
\frac{\Gamma(n \!+\! D \!-\! 1)}{n \Gamma(n \!+\! \frac{D}2)}
\Bigl(\frac{y}4 \Bigr)^n \Biggr] \Biggr\} , \qquad \label{DeltaA} \\
\lefteqn{W(y) = \frac{H^{D-2}}{(4\pi)^{\frac{D}2}} \Biggl\{
\Gamma\Bigl(\frac{D}2 \!-\!1\Bigr)
\Bigl(\frac{4}{y}\Bigr)^{\frac{D}2 -1} \!+\! \frac{\Gamma(\frac{D}2
\!+\! 2)}{(\frac{D}2 \!-\! 2) (\frac{D}2 \!-\!1)}
\Bigl(\frac{4}{y} \Bigr)^{\frac{D}2-2} } \nonumber \\
& & \hspace{3cm} \!+\! \frac{\Gamma(\frac{D}2 \!+\! 3)}{2 (\frac{D}2
\!-\! 3) (\frac{D}2\!-\!2)} \Bigl(\frac{4}{y} \Bigr)^{\frac{D}2-3}
\!+\! W_1 \!+\!
W_2 \Bigl(\frac{y \!-\!2}4\Bigr) \nonumber \\
& & \hspace{1.5cm} + \sum_{n=2}^{\infty} \Biggl[
\frac{\Gamma(n\!+\!\frac{D}2\!+\!2)
(\frac{y}4)^{n-\frac{D}2+2}}{(n\!-\! \frac{D}2\!+\!2) (n \!-\!
\frac{D}2 \!+\!1) (n \!+\! 1)!} - \frac{\Gamma(n \!+\! D)
(\frac{y}4)^n }{n (n \!-\!1) \Gamma(n \!+\! \frac{D}2)} \Biggr]
\Biggr\} , \qquad \label{DeltaW}
\end{eqnarray}
And the $D$-depdendent constants $A_1$, $W_1$ and $W_2$ are,
\begin{eqnarray}
A_1 & = & \frac{\Gamma(D\!-\!1)}{\Gamma(\frac{D}2)} \Biggl\{
-\psi\Bigl(1 \!-\! \frac{D}2\Bigr) + \psi\Bigl(\frac{D\!-\!1}2\Bigr)
+
\psi(D \!-\!1) + \psi(1) \Biggr\} , \\
W_1 & = & \frac{\Gamma(D\!+\!1)}{\Gamma(\frac{D}2 \!+\!1)} \Biggl\{
\frac{D \!+\!1}{2 D} \Biggr\} , \\
W_2 & = & \frac{\Gamma(D\!+\!1)}{\Gamma(\frac{D}2 \!+\!1)} \Biggl\{
\psi\Bigl(-\frac{D}2\Bigr) - \psi\Bigl(\frac{D\!+\!1}2\Bigr) -
\psi(D \!+\!1) - \psi(1) \Biggr\} .
\end{eqnarray}

A problem we shall often encounter consists of integrating one
propagator against another. The result can be represented as the
solution of a modified propagator equation,
\begin{equation}
\Bigl[ \square + b^2 H^2 - \Bigl(\frac{D \!-\! 1}2\Bigr)^2
H^2 \Bigr] i \Delta_{bc}(x;z) = i\Delta_c(x;z) \; . \label{Int1}
\end{equation}
The solution is easily seen to be \cite{MTW1,MTW2},
\begin{equation}
i\Delta_{bc}(x;z) = \frac1{(b^2 \!-\! c^2) H^2} \Bigl[
i\Delta_c(x;z) \!-\! i\Delta_b(x;z)\Bigr] = i\Delta_{cb}(x;z) \; .
\label{Int2}
\end{equation}
For the special case that the indices $b$ and $c$ agree one gets a
derivative,
\begin{equation}
i\Delta_{bb}(x;z) = -\frac1{2 b H^2} \frac{\partial}{\partial b}
i\Delta_b(x;z) \; . \label{Int3}
\end{equation}

We can obviously continue the process {\it ad infinitum}. For
example, consider the case where the source is not a propagator but
rather a singly integrated propagator,
\begin{equation}
\Bigl[ \square + b^2 H^2 - \Bigl(\frac{D \!-\! 1}2\Bigr)^2 H^2
\Bigr] i \Delta_{bcd}(x;z) = i\Delta_{cd}(x;z) \; . \label{Int4}
\end{equation}
The solution can be written in a form which is manifestly symmetric
under any interchange of the three indices $a$, $b$ and $c$,
\begin{eqnarray}
i\Delta_{bcd}(x;z) & = & \frac{ i\Delta_{bd}(x;z) \!-\!
i\Delta_{bc}(x;z)}{(c^2 \!-\! d^2) H^2} \; , \qquad \label{Int5a} \\
& = & \frac{ (d^2 \!-\! c^2) i\Delta_{b}(x;z) \!+\! (b^2 \!-\! d^2)
i\Delta_{c}(x;z) \!+\! (c^2 \!-\! b^2) i\Delta_{d}(x;z) }{(b^2 \!-\!
c^2) (c^2 \!-\! d^2) (d^2 \!-\! b^2) H^4} \; . \qquad \label{Int5b}
\end{eqnarray}
The case in which two of the indices are the same gives,
\begin{equation}
i\Delta_{bcc}(x;z) = -\frac1{2 c H^2} \frac{\partial}{\partial c}
i\Delta_{bc}(x;z) = \frac{ i\Delta_{cc}(x;z) \!-\!
i\Delta_{bc}(x;z)}{ (b^2 \!-\! c^2) H^2} \; . \label{Int6}
\end{equation}
And equating all three indices produces,
\begin{eqnarray}
i\Delta_{bbb}(x;z) & = & -\frac1{2 b H^2} \frac{\partial}{\partial
b} i\Delta_{bc}(x;z) \Bigl\vert_{c = b} \; , \qquad \label{Int7a} \\
& = & -\frac1{8 b^3 H^4} \Biggl[ \frac{\partial}{\partial b}
i\Delta_b(x;z) \!-\! b \Bigl(\frac{\partial}{\partial b}\Bigr)^2
i\Delta_b(x;z) \Biggr] \; . \label{Int7b}
\end{eqnarray}

\section{Vector Propagators}

One can see from (\ref{vectorL}) that the vector propagator obeys
the equation,
\begin{equation}
\Bigl[ \square \!-\! (D \!-\! 1) H^2 \!-\! M_V^2 \Bigr] i
\Bigl[\mbox{}_{\mu} \Delta_{\rho}\Bigr](x;z) = \frac{i g_{\mu\rho}
\delta^D( x\!-\! z)}{\sqrt{-g}} \; . \label{veceqn}
\end{equation}
Note that we do not assume transversality; indeed, the full vector
propagator cannot be transverse because the right hand side of
equation (\ref{veceqn}) is not transverse. The first part of this
section describes how to decompose the full propagator into its
transverse and longitudinal parts, {\it without} making any
assumptions about its eventual de Sitter invariance. Our technique
is to express these parts using projectors formed from covariant
derivative operators, acting on scalar structure functions. In the
second part we derive a scalar equation for the longitudinal
structure function and solve it using the techniques of section 3.
In the final part we carry out the same analysis for the transverse
structure function. The techniques employed here are a paradigm for
the work of the subsequent section on the graviton propagator.

\subsection{Transverse and Logitudinal Parts}

The full vector propagator can be written as the sum of a transverse
part and a longitudinal part,
\begin{equation}
i\Bigl[\mbox{}_{\mu} \Delta_{\rho}\Bigr](x;z) = i\Bigl[\mbox{}_{\mu}
\Delta^T_{\rho}\Bigr](x;z) + i\Bigl[\mbox{}_{\mu}
\Delta^L_{\rho}\Bigr](x;z) \; . \label{decomp}
\end{equation}
In previous studies \cite{AJ,TW2} the vector propagator was
expressed as a linear combination of de Sitter invariant basis
tensors like those introduced at the end of section 2. Then the
coefficient functions were chosen to enforce transversality (or
longitudinality). This method is not open to us because we cannot
assume de Sitter invariance for general mass $M_V$. What we require
instead is a covariant decomposition which entails no assumption
about de Sitter invariance.

The longitudinal part is easy,
\begin{equation}
i\Bigl[\mbox{}_{\mu} \Delta^L_{\rho}\Bigr](x;z) \equiv
\frac{\partial}{\partial x^{\mu}} \frac{\partial}{\partial z^{\rho}}
\Bigl[\mathcal{S}_{L}(x;z) \Bigr] \; . \label{longform}
\end{equation}
This expression is longitudinal for any choice of the longitudinal
structure function $\mathcal{F}_{L}(x;z)$. After much consideration
we decided to express the transverse part as,
\begin{equation}
i\Bigl[\mbox{}_{\mu} \Delta^T_{\rho}\Bigr](x;z) =
\mathcal{P}^{\alpha\beta}_{\mu}(x) \times
\mathcal{P}^{\kappa\lambda}_{\rho}(z) \times
\mathcal{Q}_{\alpha\kappa}(x;z) \times \Bigl[ \mathcal{R}_{\beta
\lambda}(x;y) \; \mathcal{S}_{T}(x;z)\Bigr] \; . \label{transform}
\end{equation}
These symbols require explanation. The differential operator
$\mathcal{P}^{\alpha\beta}_{\mu}$ is defined by writing the Maxwell
field strength tensor as $F^{\alpha\beta} =
\mathcal{P}^{\alpha\beta}_{\mu} A^{\mu}$,
\begin{equation}
\mathcal{P}^{\alpha\beta}_{\mu} \equiv \delta^{\beta}_{\mu}
D^{\alpha} \!-\! \delta^{\alpha}_{\mu} D^{\beta} \; .
\end{equation}
Note that acting $\mathcal{P}^{\alpha\beta}_{\mu}(x) \times
\mathcal{P}^{\kappa\lambda}_{\rho}(z)$ on any 4-index, symmetric
function of $x$ and $z$ produces something with the right properties
to be a transverse propagator. Of course some choices for the
4-index function give simpler final results than others! The best
selection seems to be taking two of the four indices to be more
covariant derivatives and the other two to belong to the section 2
basis tensor (\ref{dydxdx'}) which gives a Lorentz metric in the
flat space limit. This corresponds to form (\ref{transform}) with
the definitions,
\begin{eqnarray}
\mathcal{Q}_{\alpha\kappa}(x;z) & \equiv & -\frac1{2 H^2}
\frac{D}{D x^{\alpha}} \frac{D}{D z^{\kappa}} \; , \label{Qdef} \\
\mathcal{R}_{\beta\lambda}(x;z) & \equiv & -\frac1{2 H^2}
\frac{\partial^2 y(x;z)}{\partial x^{\beta} \partial z^{\lambda}} \;
. \label{Rdef}
\end{eqnarray}

\subsection{Solution for the Longitudinal Part}

To derive an equation for the longitudinal structure function we
take the divergence of the full propagator equation (\ref{veceqn}),
substitute relations (\ref{decomp}), (\ref{longform}) and
(\ref{transform}), and then commute the derivative to the left,
\begin{eqnarray}
\lefteqn{D_z^{\rho} \Bigl[\square_x \!-\! (D\!-\!1) H^2 \!-\!
M_V^2\Bigr] i\Bigl[\mbox{}_{\mu}\Delta_{\rho}\Bigr](x;z) } \nonumber
\\
& & \hspace{4cm} = \Bigl[\square_x \!-\! (D\!-\!1) H^2 \!-\!
M_V^2\Bigr] \frac{\partial}{\partial
x^{\mu}} \square_z \mathcal{S}_{L}(x;z) \; , \qquad \\
& & \hspace{4cm} = D_{\mu}^x \Bigl[\square_x \!-\! M_V^2\Bigr]
\square_z \mathcal{S}_{L}(x;z) \; , \qquad \\
& & \hspace{4cm} = -D_{\mu}^x \Biggl( \frac{i \delta^D(x \!-\!
z)}{\sqrt{-g}} \Biggr) \; .
\end{eqnarray}
Hence we conclude,
\begin{equation}
\Bigl[\square_x \!-\! M_V^2\Bigr] \square_z \mathcal{S}_{L}(x;z) =
-\frac{i\delta^D(x \!-\! z)}{\sqrt{-g}} \; .
\end{equation}
From relation (\ref{bprop}) of section 3 this implies,
\begin{equation}
\square_z \mathcal{S}_{L}(x;z) = -i\Delta_b(x;z) \qquad {\rm for}
\qquad b^2 = \Bigl(\frac{D\!-\!1}2\Bigr)^2 - \frac{M_V^2}{H^2} \; .
\end{equation}
The final solution for $\mathcal{S}_{L}$ follows from relations
(\ref{Int1}-\ref{Int2}),
\begin{equation}
\mathcal{S}_{L}(x;z) = \frac1{M_V^2} \Bigl[ i\Delta_A(x;z) \!-\!
i\Delta_b(x;z)\Bigr] = -i\Delta_{Ab}(x;z) \; . \label{longsol}
\end{equation}
We remind the reader of special case $A$ with index $b_A =
(\frac{D-1}2)$ and the explicit expansion for $i\Delta_{A}(x;z)$
given by expressions (\ref{Adef}), (\ref{dsbA}) and (\ref{DeltaA}).

\subsection{Solution for the Transverse Part}

Substituting our explicit solution (\ref{longsol}) for the
longitudinal structure function into the full propagator equation
(\ref{veceqn}) allows us to derive an equation for the transverse
part that was previously guessed \cite{TW2},
\begin{equation}
\Bigl[ \square \!-\! (D \!-\! 1) H^2 \!-\! M_V^2 \Bigr] i
\Bigl[\mbox{}_{\mu} \Delta^T_{\rho}\Bigr](x;z) = \frac{i g_{\mu\rho}
\delta^D( x\!-\! z)}{\sqrt{-g}} + \frac{\partial}{\partial x^{\mu}}
\frac{\partial}{\partial z^{\rho}} i\Delta_A(x;z) \; .
\label{transeqn}
\end{equation}
The most effective technique for solving this equation is to reduce
each side of the equation to the standard transverse form
(\ref{transform}). We then read off a scalar equation for the
transverse structure function $\mathcal{S}_{T}(x;z)$, which can be
solved by the methods of section 3.

It is best to begin by establishing some important properties of the
quadratic differential operator,
\begin{equation}
\mathbf{P}_{\mu}^{~\beta} \equiv \mathcal{P}^{\alpha\beta}_{\mu}
\times D_{\alpha} = \delta_{\mu}^{~\beta} \square - D^{\beta}
D_{\mu} \; . \label{bfP1}
\end{equation}
We shall always contract $\mathbf{P}_{\mu}^{~\beta}$ into some
vector $T_{\beta}$, so it is possible to commute the final covariant
derivatives to reach the form,
\begin{equation}
\mathbf{P}_{\mu}^{~\beta} T_{\beta} = \Biggl( \delta_{\mu}^{~\beta}
\Bigl[\square \!-\! (D\!-\! 1)H^2\Bigr] - D_{\mu} D^{\beta}\Biggr)
T_{\beta} \; . \label{commuted}
\end{equation}
It is tedious but straightforward to show that the covariant
d'Alembertian commutes with $\mathbf{P}_{\mu}^{~\beta}$,
\begin{equation}
\square \mathbf{P}_{\mu}^{~\beta} T_{\beta} =
\mathbf{P}_{\mu}^{~\beta} \square T_{\beta} \; . \label{dalcom}
\end{equation}
Note also that $\mathbf{P}_{\mu}^{~\beta}$ is transverse on both
left and right,
\begin{equation}
D^{\mu} \mathbf{P}_{\mu}^{~\beta} T_{\beta} = 0 =
\mathbf{P}_{\mu}^{~\beta} D_{\beta} T \; . \label{doubleT}
\end{equation}
$\mathbf{P}_{\mu}^{~\beta}$ must therefore be proportional to
transverse projection operator. The proportionality factor can be
found by squaring. Comparing relations (\ref{doubleT}) and
(\ref{commuted}) implies,
\begin{equation}
\mathbf{P}_{\mu}^{~\alpha} \times \mathbf{P}_{\alpha}^{~\beta}
T_{\beta} = \Bigl[ \square \!-\! (D\!-\!1) H^2\Bigr]
\mathbf{P}_{\mu}^{~\beta} T_{\beta} = \mathbf{P}_{\mu}^{~\beta}
\Bigl[\square \!-\! (D\!-\!1) H^2 \Bigr] T_{\beta} \; .
\label{Psquare}
\end{equation}

The relevance of $\mathbf{P}_{\mu}^{~\beta}$ is that it gives the
differential operators in front of the general transverse form
(\ref{transform}),
\begin{equation}
\mathcal{P}^{\alpha\beta}_{\mu}(x) \times
\mathcal{P}^{\kappa\lambda}_{\rho}(z) \times
\mathcal{Q}_{\alpha\kappa}(x;z) = -\frac1{2 H^2}
\mathbf{P}_{\mu}^{~\beta}(x) \times \mathbf{P}_{\rho}^{~\lambda}(z)
\; . \label{Prel}
\end{equation}
Substituting (\ref{transform}) into equation (\ref{transeqn}), and
making use or relations (\ref{Prel}) and (\ref{dalcom}), implies,
\begin{eqnarray}
\lefteqn{\Bigl[ \square \!-\! (D \!-\! 1) H^2 \!-\! M_V^2 \Bigr] i
\Bigl[\mbox{}_{\mu} \Delta^T_{\rho}\Bigr](x;z) } \nonumber \\
& & \hspace{.5cm} = \mathcal{P}^{\alpha\beta}_{\mu}(x) \times
\mathcal{P}^{\kappa\lambda}_{\rho}(z) \times
\mathcal{Q}_{\alpha\kappa}(x;z) \Bigl[ \square \!-\! (D \!-\! 1)H^2
\!-\! M_V^2\Bigr] \Bigl[ \mathcal{R}_{\beta\lambda} \mathcal{S}_T
\Bigr] \; . \qquad
\end{eqnarray}
We need next to consider what the d'Alembertian gives when acting on
the factors to the far right,
\begin{equation}
\square_x \Bigl[ \mathcal{R}_{\beta \lambda} \mathcal{S}_T \Bigr] =
\Bigl[ \square_x \mathcal{R}_{\beta\lambda} \Bigr] \mathcal{S}_T + 2
g^{\alpha\gamma} \frac{D \mathcal{R}_{\beta\lambda}}{D x^{\alpha}}
\frac{\partial \mathcal{S}_T}{\partial x^{\gamma}} +
\mathcal{R}_{\beta\lambda} \square_x \mathcal{S}_T \; . \label{actD}
\end{equation}
Recalling the definition (\ref{Rdef}) of
$\mathcal{R}_{\beta\lambda}(x;z)$, and making use of relation
(\ref{covdiv}) from section 2, gives two identities we shall use in
this section and the next,
\begin{eqnarray}
\frac{D}{D x^{\alpha}} \mathcal{R}_{\beta\lambda}(x;z) & = & \frac12
\, g_{\alpha\beta}(x) \frac{\partial y}{\partial z^{\lambda}} \; ,
\qquad \label{useful1} \\
\square \mathcal{R}_{\beta\lambda}(x;z) & = & -H^2
\mathcal{R}_{\beta\lambda}(x;z) \; . \label{useful2}
\end{eqnarray}
Substitute these in (\ref{actD}) and pass the single derivative back
outside to obtain,
\begin{eqnarray}
\square_x \Bigl[ \mathcal{R}_{\beta\lambda} \mathcal{S}_T \Bigr] & =
& \frac{\partial y}{\partial z^{\lambda}} \frac{\partial
\mathcal{S}_T}{\partial x^{\beta}} + \mathcal{R}_{\beta\lambda}
\Bigl[ \square_x \!-\! H^2\Bigr] \mathcal{S}_T \; , \qquad \\
& = & \frac{\partial}{\partial x^{\beta}} \Bigl[ \frac{\partial
y}{\partial z^{\lambda}} \mathcal{S}_T \Bigr] - \frac{\partial^2
y}{\partial x^{\beta} \partial z^{\lambda}} \mathcal{S}_T +
\mathcal{R}_{\beta\lambda} \Bigl[\square_x \!-\! H^2\Bigr]
\mathcal{S}_T \; , \qquad \\
& = & \frac{\partial}{\partial x^{\beta}} \Bigl[ \frac{\partial
y}{\partial z^{\lambda}} \mathcal{S}_T \Bigr] + \mathcal{R}_{\beta
\lambda} \Bigl[ \square_x \!+\! H^2\Bigr] \mathcal{S}_T \; . \qquad
\label{fullpass}
\end{eqnarray}
The first term on the right of equation (\ref{fullpass}) is
longitudinal. In view of relation (\ref{doubleT}) we therefore
conclude,
\begin{eqnarray}
\lefteqn{\Bigl[ \square \!-\! (D \!-\! 1) H^2 \!-\! M_V^2 \Bigr] i
\Bigl[\mbox{}_{\mu} \Delta^T_{\rho}\Bigr](x;z) } \nonumber \\
& & \hspace{.5cm} = \mathcal{P}^{\alpha\beta}_{\mu}(x) \times
\mathcal{P}^{\kappa\lambda}_{\rho}(z) \times
\mathcal{Q}_{\alpha\kappa}(x;z) \Biggl[ \mathcal{R}_{\beta\lambda}
\Bigl[ \square \!-\! (D \!-\! 2)H^2 \!-\! M_V^2\Bigr] \mathcal{S}_T
\Biggr] . \qquad \label{finalleft}
\end{eqnarray}

It remains to reduce the right hand side of (\ref{transeqn}) to the
standard form (\ref{transform}) we have adopted for transverse
bi-tensors,
\begin{eqnarray}
i\Bigl[\mbox{}_{\mu} P_{\rho}\Bigr](x;z) & \equiv & \frac{i
g_{\mu\rho} \delta^D( x\!-\! z)}{\sqrt{-g}} +
\frac{\partial}{\partial x^{\mu}} \frac{\partial}{\partial z^{\rho}}
i\Delta_A(x;z) \; , \label{form1} \qquad \\
& = & \mathcal{P}^{\alpha\beta}_{\mu}(x) \times
\mathcal{P}^{\kappa\lambda}_{\rho}(z) \times \mathcal{Q}_{\alpha
\kappa}(x;z) \Bigl[ \mathcal{R}_{\beta\lambda}(x;z)
\mathcal{P}_1(x;z) \Bigr] \; . \qquad \label{form2}
\end{eqnarray}
This is easily accomplished by acting $\mathbf{P}_{\nu}^{~\mu}(x)
\times \mathbf{P}_{\sigma}^{~\rho}(z)$ on both forms. Acting this
operator on (\ref{form1}) and making use of relation (\ref{doubleT})
gives,
\begin{eqnarray}
\lefteqn{\mathbf{P}_{\nu}^{~\mu}(x) \times
\mathbf{P}_{\sigma}^{~\rho}(z) \, i\Bigl[ \mbox{}_{\mu}
P_{\rho}\Bigr](x;z) } \nonumber \\
& & \hspace{1cm} = -2 H^2 \mathcal{P}^{\alpha\beta}_{\nu}(x) \times
\mathcal{P}^{\kappa\lambda}_{\sigma}(z) \times
\mathcal{Q}_{\alpha\kappa}(x;z) \Bigl[ g_{\beta\lambda} \frac{i
\delta^D(x \!-\! z)}{\sqrt{-g}} \Bigr] \; , \qquad \\
& & \hspace{1cm} = -2 H^2 \mathcal{P}^{\alpha\beta}_{\nu}(x) \times
\mathcal{P}^{\kappa\lambda}_{\sigma}(z) \times
\mathcal{Q}_{\alpha\kappa}(x;z) \Bigl[
\mathcal{R}_{\beta\lambda}(x;z) \frac{i \delta^D(x \!-\!
z)}{\sqrt{-g}} \Bigr] \; . \qquad \label{exp1}
\end{eqnarray}
Acting instead on (\ref{form2}) and making use of relations
(\ref{Psquare}) and (\ref{fullpass}) gives,
\begin{eqnarray}
\lefteqn{ \mathbf{P}_{\nu}^{~\mu}(x) \times
\mathbf{P}_{\sigma}^{~\rho}(z) \, i\Bigl[ \mbox{}_{\mu}
P_{\rho}\Bigr](x;z) } \nonumber \\
& & \hspace{1cm} = \mathcal{P}^{\alpha\beta}_{\nu}(x) \times
\mathcal{P}^{\kappa\lambda}_{\sigma}(z) \times
\mathcal{Q}_{\alpha\kappa}(x;z) \Bigl[ \square \!-\! (D \!-\! 1)
H^2\Bigr]^2 \Bigl[ \mathcal{R}_{\beta\lambda} \mathcal{P}_1 \Bigr]
\; , \qquad \\
& & \hspace{1cm} = \mathcal{P}^{\alpha\beta}_{\nu}(x) \times
\mathcal{P}^{\kappa\lambda}_{\sigma}(z) \times
\mathcal{Q}_{\alpha\kappa}(x;z) \Biggl[ \mathcal{R}_{\beta\lambda}
\Bigl[ \square \!-\! (D \!-\! 2) H^2\Bigr]^2 \mathcal{P}_1 \Biggr]
\; . \qquad \label{exp2}
\end{eqnarray}
Comparing expressions (\ref{exp1}) and (\ref{exp2}) implies,
\begin{equation}
\Bigl[ \square \!-\! (D\!-\!2) H^2\Bigr]^2 \mathcal{P}_1(x;z) = -2
H^2 \frac{i \delta^D(x \!-\! z)}{\sqrt{-g}} \; .
\end{equation}
Relation (\ref{bprop}) from section 3 --- with the special case of
$b = (\frac{D-3}2)$ --- to infer,
\begin{equation}
\Bigl[\square \!-\! (D\!-\!2)H^2\Bigr] \mathcal{P}_1(x;z) = -2 H^2
i\Delta_B(x;z) \; .
\end{equation}
Now apply relations (\ref{Int1}-\ref{Int3}) to finally obtain the
structure function for the transverse projection functional,
\begin{equation}
\mathcal{P}_1(x;z) = -2 H^2 i\Delta_{BB}(x;z) \; .
\end{equation}

We have now reduced the transverse propagator equation to the form,
\begin{eqnarray}
\lefteqn{\mathcal{P}^{\alpha\beta}_{\mu}(x) \times
\mathcal{P}^{\kappa\lambda}_{\rho}(z) \times
\mathcal{Q}_{\alpha\kappa}(x;z) \Biggl[ \mathcal{R}_{\beta\lambda}
\Bigl[ \square \!-\! (D \!-\! 2)H^2 \!-\! M_V^2\Bigr] \mathcal{S}_T
\Biggr] } \nonumber \\
& & \hspace{2.5cm} = \mathcal{P}^{\alpha\beta}_{\mu}(x) \times
\mathcal{P}^{\kappa\lambda}_{\rho}(z) \times
\mathcal{Q}_{\alpha\kappa}(x;z) \Biggl[ \mathcal{R}_{\beta\lambda}
\Bigl[-2 H^2 i\Delta_{BB}\Bigr] \Biggr] \; . \qquad
\end{eqnarray}
The transverse structure function therefore obeys,
\begin{equation}
\Bigl[\square \!-\! (D \!-\! 2) H^2 \!-\! M_V^2 \Bigr] \mathcal{S}_T
= -2H^2 i\Delta_{BB}(x;z) \; .
\end{equation}
Again making use of relations (\ref{Int1}-\ref{Int3}), our solution
for it is,
\begin{equation}
\mathcal{S}_T = + \frac{2 H^2}{M_V^2} i\Delta_{BB} + \frac{2
H^2}{M_V^4} \Bigl[ i\Delta_B - i\Delta_c\Bigr] \quad {\rm where}
\quad c = \sqrt{ \Bigl( \frac{D \!-\! 3}2\Bigr)^2 -
\frac{M_V^2}{H^2} } \; . \label{transsol}
\end{equation}

\section{The Graviton Propagator}

The previous section provides a model for the analysis of this
section, except that we immediately specialize to gravitons which
obey de Donder gauge,
\begin{equation}
D^{\mu} h_{\mu\nu} - \frac12 D_{\nu} h^{\mu}_{~\mu} = 0 \; .
\label{deDonder}
\end{equation}
The first task is to express the propagator of such a graviton in
terms of covariant projectors acting on scalar structure functions.
With just a small extension of our previous results we can then
commute the differential operator to act directly on the structure
functions. The final step is identifying the de Donder gauge
projection functionals.

\subsection{Enforcing de Donder Gauge}

In de Donder gauge (\ref{deDonder}) the propagator must obey the
gauge condition on either coordinate and its associated index group,
\begin{eqnarray}
\Bigl[ \delta^{\alpha}_{\mu} D_x^{\beta} - \frac12 D^x_{\mu}
g^{\alpha\beta}(x)\Bigr] \times i \Bigl[\mbox{}_{\alpha\beta}
\Delta_{\rho\sigma} \Bigr](x;z) & = & 0 \; , \label{dD1} \\
\Bigl[ \delta^{\alpha}_{\rho} D_z^{\beta} - \frac12 D^z_{\rho}
g^{\alpha\beta}(z)\Bigr] \times i \Bigl[\mbox{}_{\mu\nu}
\Delta_{\alpha\beta} \Bigr](x;z) & = & 0 \; . \label{dD2}
\end{eqnarray}
Just as was the case for the vector propagator with the analogous
conditions of transversality and longitudinality, we seek here to
enforce (\ref{dD1}-\ref{dD2}) by acting covariant projectors on
scalar structure functions. It turns out there are two ways to do
it, corresponding to the spin zero and spin two parts of the
graviton propagator,
\begin{equation}
i \Bigl[\mbox{}_{\alpha\beta} \Delta_{\rho\sigma} \Bigr](x;z) = i
\Bigl[\mbox{}_{\alpha\beta} \Delta^0_{\rho\sigma} \Bigr](x;z) + i
\Bigl[\mbox{}_{\alpha\beta} \Delta^2_{\rho\sigma} \Bigr](x;z) \; .
\label{gravdecomp}
\end{equation}

The spin zero part of the graviton propagator is almost as simple as
the longitudinal part of the vector propagator. It is a linear
combination of longitudinal and trace terms from each index group,
\begin{equation}
i\Bigl[\mbox{}_{\mu\nu} \Delta^0_{\rho\sigma}\Bigr](x;z) =
\mathcal{P}_{\mu\nu}(x) \times \mathcal{P}_{\rho\sigma}(z)
\Bigl[\mathcal{S}_0(x;z) \Bigr] \; . \label{Spin0}
\end{equation}
The projector $\mathcal{P}_{\mu\nu}$ is,
\begin{equation}
\mathcal{P}_{\mu\nu} \equiv D_{\mu} D_{\nu} + \frac{g_{\mu\nu}}{D \!-\!2}
\Bigl[ \square \!+\! 2 (D \!-\! 1) H^2 \Bigr] \; . \label{P0op}
\end{equation}

Unlike the spin zero part of the graviton propagator, the spin two
part is both transverse and also traceless within each index group.
Recall that we obtained the key projector for the transverse part of
the photon propagator by writing the Maxwell field strength tensor
as $F^{\alpha\beta} = \mathcal{P}^{\alpha\beta}_{\mu} A^{\mu}$. We
similarly define the projector
$\mathcal{P}^{\alpha\beta\gamma\delta}_{\mu\nu}$ by expanding the
Weyl tensor in powers of the graviton field
$C^{\alpha\beta\gamma\delta} =
\mathcal{P}^{\alpha\beta\gamma\delta}_{\mu\nu} h^{\mu\nu} + O(h^2)$.
The final result takes the form \cite{PW},
\begin{eqnarray}
\lefteqn{\mathcal{P}_{\mu\nu}^{\alpha\beta\gamma\delta} \equiv
\mathcal{D}_{\mu\nu}^{\alpha\beta\gamma\delta} + \frac1{D \!-\!2}
\Bigl[ g^{\alpha\delta} \mathcal{D}^{\beta\gamma}_{\mu\nu} \!-\!
g^{\beta\delta} \mathcal{D}^{\alpha\gamma}_{\mu\nu} \!-\!
g^{\alpha\gamma} \mathcal{D}^{\beta\delta}_{\mu\nu} \!+\!
g^{\beta\gamma} \mathcal{D}^{\alpha\delta}_{\mu\nu} \Bigr] } \nonumber \\
& & \hspace{5cm} + \frac1{(D \!-\! 1) (D \!-\! 2)} \Bigl[ g^{\alpha
\gamma} g^{\beta \delta} \!-\! g^{\alpha\delta} g^{\beta \gamma}
\Bigr] \mathcal{D}_{\mu\nu} \; . \qquad \label{CPdef}
\end{eqnarray}
The various pieces of this are,
\begin{eqnarray}
\mathcal{D}^{\alpha\beta\gamma\delta}_{\mu\nu} &\equiv & \frac12 \Bigl[
\delta^{\alpha}_{(\mu} \delta^{\delta}_{\nu)} D^{\gamma} D^{\beta} \!-\!
\delta^{\beta}_{(\mu} \delta^{\delta}_{\nu)} D^{\gamma} D^{\alpha} \!-\!
\delta^{\alpha}_{(\mu} \delta^{\gamma}_{\nu)} D^{\delta} D^{\beta} \!+\!
\delta^{\beta}_{(\mu} \delta^{\gamma}_{\nu)} D^{\delta} D^{\alpha} \Bigr]
\; , \qquad \label{CDdef1} \\
\mathcal{D}^{\beta\delta}_{\mu\nu} & \equiv & g_{\alpha\gamma}
\mathcal{D}^{\alpha\beta\gamma\delta}_{\mu\nu} = \frac12 \Bigl[
\delta^{\delta}_{(\mu} D_{\nu)} D^{\beta} \!-\! \delta^{\beta}_{(\mu}
\delta^{\delta}_{\nu)} \square \!-\! g_{\mu\nu} D^{\delta} D^{\beta}
\!+\! \delta^{\beta}_{(\mu} D^{\delta} D_{\nu)} \Bigr] \; , \label{CDdef2} \\
\mathcal{D}_{\mu\nu} & \equiv & g_{\alpha\gamma} g_{\beta\delta}
\mathcal{D}^{\alpha\beta\gamma\delta}_{\mu\nu} = D_{(\mu} D_{\nu)} -
g_{\mu\nu} \square \; . \label{CDdef3}
\end{eqnarray}
Acting $\mathcal{P}^{\alpha\beta\gamma\delta}_{\mu\nu}(x) \times
\mathcal{P}^{\kappa\lambda\theta\phi}_{\rho\sigma}(z)$ on any eight
index, symmetric function of $x$ and $z$ would produce a transverse
and traceless tensor but, as with the vector, it pays to select a
simple form. The best choice seems to be taking half the indices in
the form of more covariant derivative operators, and the other half
from two factors of the mixed second derivative (\ref{dydxdx'}) of
the length function,
\begin{equation}
i\Bigl[\mbox{}_{\mu\nu} \Delta^2_{\rho\sigma}\Bigr](x;z) =
\mathcal{P}_{\mu\nu}^{\alpha\beta\gamma\delta}(x) \times
\mathcal{P}_{\rho\sigma}^{\kappa\lambda\theta\phi}(z) \times
\mathcal{Q}_{\alpha\kappa} \times \mathcal{Q}_{\gamma\theta} \Bigl[
\mathcal{R}_{\beta\lambda} \mathcal{R}_{\delta\phi} \,
\mathcal{S}_2(x;z) \Bigr] \; . \label{Spin2}
\end{equation}
We remind the reader of the definitions (\ref{Qdef}-\ref{Rdef}) of
$\mathcal{Q}_{\alpha\kappa}(x;z)$ and
$\mathcal{R}_{\beta\lambda}(x;z)$.

We close this subsection by giving the propagator equation. Acting
the Lichnerowicz operator (\ref{Lichnerowicz}) on the graviton field
and making use of the de Donder gauge condition (\ref{deDonder})
gives,
\begin{equation}
-\mathbf{D}^{\mu\nu\rho\sigma} h_{\rho\sigma} = \frac12 \Bigl[
\square \!-\! 2 H^2\Bigr] h^{\mu\nu} - \frac14 g^{\mu\nu} \Bigl[
\square \!+\! 2 (D \!-\! 3) H^2 \Bigr] h^{\rho}_{~\rho} \; .
\end{equation}
This means the propagator obeys a relation of the form,
\begin{eqnarray}
\lefteqn{\frac12 \Bigl[ \square_x \!-\! 2 H^2\Bigr]
i\Bigl[\mbox{}_{\mu\nu} \Delta_{\rho\sigma}\Bigr](x;z) - \frac14
g_{\mu\nu}(x) \Bigl[ \square_x \!+\! 2(D \!-\!3) H^2\Bigr] i\Bigl[
\mbox{}^{\alpha}_{~\alpha} \Delta_{\rho\sigma}\Bigr](x;z) }
\nonumber \\
& & \hspace{4.5cm} =\! g_{\mu (\rho} g_{\sigma) \nu} \times
\frac{i\delta^D(x \!-\!z)}{\sqrt{-g}} + \Bigl( {\rm Other\ Terms}
\Bigr) , \qquad \label{propeqn1}
\end{eqnarray}
where the ``Other Terms'' make the right hand side consistent with
the gauge condition. However, the right hand side of
(\ref{propeqn1}) cannot be symmetric under the interchange of
$x^{\mu}$ and $z^{\mu}$ (and their associated indices) because the
left hand side of the equation obeys de Donder gauge on $z^{\mu}$
but not on $x^{\mu}$. It is better to subtract off a term
proportional to the trace,
\begin{eqnarray}
\lefteqn{\frac12 \Bigl[ \square_x \!-\! 2 H^2\Bigr]
i\Bigl[\mbox{}_{\mu\nu} \Delta_{\rho\sigma}\Bigr](x;z) - \frac14
g_{\mu\nu}(x) \Bigl[ \square_x \!+\! 2(D \!-\!3) H^2\Bigr] i\Bigl[
\mbox{}^{\alpha}_{~\alpha} \Delta_{\rho\sigma}\Bigr](x;z) }
\nonumber \\
& & \hspace{2.5cm} -\frac{g_{\mu\nu}}{D\!-\!2} \times -\Bigl(\frac{D
\!-\! 2}4\Bigr) \Bigl[ \square \!+\! 2 (D \!-\! 1) H^2 \Bigr] \,
i\Bigl[\mbox{}^{\alpha}_{\alpha} \Delta_{\rho\sigma}\Bigr](x;z) \; ,
\qquad \\
& & \hspace{1cm} = \frac12 \Bigl[ \square_x \!-\! 2 H^2\Bigr]
i\Bigl[\mbox{}_{\mu\nu} \Delta_{\rho\sigma}\Bigr](x;z) + H^2
g_{\mu\nu}(x) i\Bigl[\mbox{}^{\alpha}_{~\alpha} \Delta_{\rho\sigma}
\Bigr](x;z) \; . \qquad \label{goodform}
\end{eqnarray}
It is easily checked that (\ref{goodform}) obeys the de Donder gauge
condition on both $x^{\mu}$ and $z^{\mu}$. Hence the right hand side
of the equation is symmetric under interchange of $x^{\mu}$ and
$z^{\mu}$ and can in fact be guessed \cite{MTW2},
\begin{eqnarray}
\lefteqn{\frac12 \Bigl[ \square_x \!-\! 2 H^2\Bigr]
i\Bigl[\mbox{}_{\mu\nu} \Delta_{\rho\sigma}\Bigr](x;z) + H^2
g_{\mu\nu}(x) i\Bigl[ \mbox{}^{\alpha}_{~\alpha}
\Delta_{\rho\sigma}\Bigr](x;z) } \nonumber \\
& & \hspace{-.7cm} =\! \Bigl[g_{\mu (\rho} g_{\sigma) \nu} \!-\!
\frac{g_{\mu\nu} g_{\rho\sigma}}{D \!-\!2} \Bigr] \frac{i\delta^D(x
\!-\!z)}{\sqrt{-g}} + \frac12 \! \left\{ \!\matrix{ D^x_{\mu}
D^z_{\rho} i [\mbox{}_{\nu} \Delta^W_{\sigma}] + D^x_{\mu}
D^z_{\sigma} i [\mbox{}_{\nu} \Delta^W_{\rho}] \cr + D^x_{\nu}
D^z_{\rho} i [\mbox{}_{\mu} \Delta^W_{\sigma}] + D^x_{\nu}
D^z_{\sigma} i [\mbox{}_{\mu} \Delta^W_{\rho}] \cr} \!\right\} .
\qquad \label{propeqn2}
\end{eqnarray}
Here $i[\mbox{}_{\mu} \Delta^W_{\rho}]$ is the full vector
propagator for the tachyonic mass $M_V^2 = -2(D-1) H^2$, which obeys
the equation,
\begin{equation}
\Bigl[ \square \!+\! (D \!-\! 1) H^2\Bigr] i \Bigl[ \mbox{}_{\mu}
\Delta^W_{\rho}\Bigr](x;z) = \frac{i g_{\mu\rho} \delta^D(x \!-\!
z)}{\sqrt{-g}} \; .
\end{equation}
Recall from section 4 that it has the form given by equations
(\ref{decomp}), (\ref{longform}) and (\ref{transform}). From
equations (\ref{longsol}) and (\ref{transsol}) we see that the
longitudinal and transverse structure functions are,
\begin{eqnarray}
\mathcal{S}_{L}(x;z) & = & -i\Delta_{AM}(x;z) \; , \qquad \label{SL} \\
\mathcal{S}_{T}(x;z) & = & \frac1{D \!-\! 1} \Bigl[
-i\Delta_{BB}(x;z) \!+\! i\Delta_{BW}(x;z)\Bigr] \; . \qquad
\label{ST}
\end{eqnarray}

\subsection{The Spin Zero Part}

To derive an equation for the spin zero structure function we simply
take the trace of the propagator equation (\ref{propeqn2}). Tracing
on the left hand side and making use of relations
(\ref{gravdecomp}-\ref{P0op}) gives,
\begin{eqnarray}
\lefteqn{\frac12 \Bigl[ \square_x \!+\! 2 (D \!-\! 1) H^2 \Bigr] \,
i\Bigl[\mbox{}^{\alpha}_{~\alpha} \Delta_{\rho\sigma}\Bigr](x;z) }
\nonumber \\
& & \hspace{1cm} = \Bigl( \frac{D \!-\! 1}{D \!-\! 2}\Bigr)
\Bigl[\square_x \!+\! 2 (D \!-\! 1) H^2\Bigr] \Bigl[ \square_x \!+\!
D H^2 \Bigr] \mathcal{P}_{\rho\sigma}(z) \Bigl[ \mathcal{S}_{0}(x;z)
\Bigr] \; . \label{Trace1} \qquad
\end{eqnarray}
Tracing the right hand side of (\ref{propeqn2}) and making use of
relations (\ref{SL}) and (\ref{Int1}-\ref{Int3}) implies,
\begin{eqnarray}
\lefteqn{\frac12 \Bigl[ \square_x \!+\! 2 (D\!-\!1) H^2 \Bigr] i
\Bigl[ \mbox{}^{\alpha}_{~\alpha} \Delta_{\rho\sigma}\Bigr](x;z) } \nonumber \\
& & \hspace{3cm} = -\frac{2}{D \!-\!2} \frac{g_{\rho\sigma} i
\delta^D(x \!-\! z)}{\sqrt{-g}} - 2 D^z_{\rho} D^z_{\sigma} i\Delta_M(x;z) \; , \qquad \\
& & \hspace{3cm} = -2 \mathcal{P}_{\rho\sigma}(z) i\Delta_{M}(x;z)
\; . \label{Trace2}
\end{eqnarray}
The equation for $\mathcal{S}_0(x;z)$ derives from comparing
expressions (\ref{Trace1}) and (\ref{Trace2}),
\begin{equation}
\Bigl[ \square \!+\! 2 (D \!-\! 1) H^2 \Bigr] \Bigl[ \square \!+\! D
H^2\Bigr] \mathcal{S}_0(x;z) = -2 \Bigl( \frac{D \!-\! 2}{D \!-\!
1}\Bigr) i\Delta_M(x;z) \; . \label{S0eqn}
\end{equation}
Its solution follows easily from relations (\ref{Int1}-\ref{Int7b}),
\begin{equation}
\mathcal{S}_0(x;z) = \frac{2 i\Delta_{MM}(x;z) \!-\! 2 i
\Delta_{MW}(x;z)}{(D \!-\! 1) H^2} = -2 \Bigl( \frac{D \!-\! 2}{D
\!-\! 1}\Bigr) i\Delta_{WMM}(x;z) \; . \label{spin0sol}
\end{equation}

\subsection{The Spin Two Part}

This is the most complicated analysis we shall have to make and it
is greatly facilitated by the analogy with what was done for the
transverse part of the vector propagator in section 4.3. Here, as
for that case, the first step is to derive an equation for the
remaining (spin two) part of the propagator by subtracting off the
part we already have. We then establish some identities for a
differential projector which comprises the exterior operators of the
spin two part (\ref{Spin2}) of the propagator. These properties
allow us to pass the d'Alembertian in the propagator equation
through to act on the spin two structure function
$\mathcal{S}_2(x;z)$. Squaring this operator also allows us to
express the right hand side of the propagator equation in the same
form (\ref{Spin2}) with a known structure function. Comparing the
two sides of the equation leads to a scalar differential equation
which can be solved by the techniques of section 3.

We derive an equation for the pure spin two part of the propagator
from the full equation (\ref{propeqn2}) by substituting the spin
zero structure function (\ref{spin0sol}), with definitions
(\ref{Spin0}-\ref{P0op}). Now move everything known to right hand
side to reach the form,
\begin{eqnarray}
\lefteqn{\frac12 \Bigl[ \square_x \!-\! 2 H^2\Bigr] \, i\Bigl[
\mbox{}_{\mu\nu} \Delta^2_{\rho\sigma}\Bigr](x;z) \equiv i \Bigl[
\mbox{}_{\mu\nu} P^2_{\rho\sigma} \Bigr](x;z) \; , }
\label{spin2eqn} \\
& & \hspace{.2cm} = \Bigl[ g_{\mu (\rho} g_{\sigma) \nu} -
\frac{g_{\mu\nu} g_{\rho\sigma}}{D \!-\! 2} \Bigr] \frac{i
\delta^D(x \!-\! z)}{\sqrt{-g}} + \Bigl( \frac{D \!-\!2}{D \!-\!
1}\Bigr) \mathcal{P}_{\mu\nu}(x) \times \mathcal{P}_{\rho\sigma}(z)
i\Delta_{WM}(x;z) \nonumber \\
& & \hspace{3cm} + \frac12 \left( \matrix{ D^x_{\mu} D^z_{\rho} i
[\mbox{}_{\nu} \Delta^{W}_{\sigma}](x;z) + D^x_{\mu} D^z_{\sigma} i
[\mbox{}_{\nu} \Delta^{W}_{\rho}](x;z) \cr + D^x_{\nu} D^z_{\rho} i
[\mbox{}_{\mu} \Delta^W_{\sigma}](x;z) + D^x_{\nu} D^z_{\sigma} i
[\mbox{}_{\mu} \Delta^W_{\rho}](x;z) \cr} \right) \; . \qquad
\label{P2}
\end{eqnarray}
It can easily be checked that the right hand side of (\ref{P2}) is
transverse and traceless on each index group. We will eventually
reduce this transverse-traceless projector to standard form,
\begin{equation}
i\Bigl[\mbox{}_{\mu\nu} P^2_{\rho\sigma}\Bigr](x;z) =
\mathcal{P}^{\alpha\beta\gamma\delta}_{\mu\nu}(x) \times
\mathcal{P}^{\kappa\lambda\theta\phi}_{\rho\sigma}(z) \times
\mathcal{Q}_{\alpha\kappa} \times \mathcal{Q}_{\gamma\theta}
\Bigl[\mathcal{R}_{\beta\lambda} \mathcal{R}_{\delta\phi}
\mathcal{P}_2\Bigr] \; . \label{standard}
\end{equation}
However, it is best to first concentrate on the left hand side of
the propagator equation (\ref{P2}).

In analogy with the transverse projector $\mathbf{P}_{\mu}^{~\beta}$
defined in equation (\ref{bfP1}), we define the transverse-traceless
projector,
\begin{equation}
\mathbf{P}_{\mu\nu}^{~~ \beta\delta} \equiv
\mathcal{P}^{\alpha\beta\gamma\delta}_{\mu\nu} \times D_{\alpha}
D_{\gamma} \; . \label{bfP2}
\end{equation}
We shall always consider this acted on a second rank tensor
$F_{\beta\delta}$. From the expressions (\ref{CPdef}-\ref{CDdef3})
which define $\mathcal{P}^{\alpha\beta\gamma\delta}_{\mu\nu}$ it is
straightforward but tedious to reach the form,
\begin{eqnarray}
\lefteqn{ \mathbf{P}_{\mu\nu}^{~~ \beta\delta} \times
F_{\beta\delta} = \frac12 \Bigl(\frac{D \!-\! 3}{D \!-\! 2}\Bigr)
\Biggl\{ D_{\mu} \square D^{\alpha} F_{\alpha \nu} + D_{\mu} \square
D^{\beta} F_{\nu\beta} - \square^2
F_{\mu \nu} } \nonumber \\
& & \hspace{0cm} - D_{\mu} D_{\nu} D^{\alpha} D^{\beta} F_{\alpha
\beta} + \frac1{D \!-\! 1} \Bigl[ D_{\mu} D_{\nu} \!-\! g_{\mu\nu}
\square\Bigr] \Bigl[ D^{\alpha} D^{\beta} F_{\alpha\beta} \!-\!
\square F_{\alpha}^{~\alpha} \Bigr]
\nonumber \\
& & \hspace{.5cm} + H^2 \Bigl[ -2 g_{\mu\nu} D^{\alpha} D^{\beta}
F_{\alpha\beta} -  g_{\mu\nu} \square F_{\alpha}^{~\alpha} -2
D_{\mu} D_{\nu} F_{\alpha}^{~\alpha} + D_{\mu} D^{\alpha} F_{\alpha
\nu} \nonumber \\
& & \hspace{2cm} + D_{\mu} D^{\beta} F_{\nu\beta} + (D \!+\! 2)
\square F_{\mu\nu} \Bigr] + H^4 \Bigl[ 2 g_{\mu\nu}
F_{\alpha}^{~\alpha} - 2 D F_{\mu\nu} \Bigr] \Biggr\} . \qquad
\label{bigresult}
\end{eqnarray}
(Note the multiplicative factor of $D-3$ which derives from the fact
that the Weyl tensor vanishes for $D=3$.) It is easy to see from
(\ref{bigresult}) that $\mathbf{P}_{\mu\nu}^{~~ \beta\delta}$ is
traceless on both the left and the right,
\begin{equation}
g^{\mu\nu} \mathbf{P}_{\mu\nu}^{~~\beta\delta} F_{\beta\delta} = 0 =
\mathbf{P}_{\mu\nu}^{~~\beta\delta} ( g_{\beta\delta} F) \; .
\label{traceless}
\end{equation}
It is also transverse on any index, both on the left and the right,
\begin{eqnarray}
D^{\mu} \Bigl( \mathbf{P}_{\mu\nu}^{~~\beta\delta} F_{\beta\delta}
\Bigr) = & 0 & = D^{\nu} \Bigl( \mathbf{P}_{\mu\nu}^{~~\beta\delta}
F_{\beta\delta} \Bigr) \; , \label{transverse1} \\
\mathbf{P}_{\mu\nu}^{~~\beta\delta} (D_{\beta} F_{\delta}) = & 0 & =
\mathbf{P}_{\mu\nu}^{~~\beta\delta} (D_{\delta} F_{\beta}) \; .
\end{eqnarray}
These two properties are very important because the only terms in
expression (\ref{bigresult}) which don't involve either divergences
or traces are,
\begin{eqnarray}
\lefteqn{\frac12 \Bigl(\frac{ D \!-\! 3}{D \!-\! 2} \Bigr) \Biggl\{
-\square^2 F_{\mu\nu} + (D \!+\!2) H^2 \square F_{\mu\nu} - 2 D H^4
F_{\mu\nu} \Biggr\} } \nonumber \\
& & \hspace{4.8cm} = -\frac12 \Bigl( \frac{D \!-\! 3}{D \!-\!
2}\Bigr) \Bigl[ \square \!-\! 2 H^2\Bigr] \Bigl[ \square \!-\! D
H^2\Bigr] F_{\mu\nu} \; . \qquad
\end{eqnarray}
Hence squaring $\mathbf{P}_{\mu\nu}^{~~\beta\delta}$ gives,
\begin{eqnarray}
\mathbf{P}_{\mu\nu}^{~~\alpha\gamma} \times
\mathbf{P}_{\alpha\gamma}^{~~\beta\delta} F_{\beta\delta} & = &
-\frac12 \Bigl( \frac{D \!-\! 3}{D \!-\! 2}\Bigr) \Bigl[ \square
\!-\! 2 H^2\Bigr] \Bigl[ \square \!-\! D H^2\Bigr]
\mathbf{P}_{\mu\nu}^{~~\beta\delta} F_{\beta\delta} \; , \qquad \\
& = & -\frac12 \Bigl( \frac{D \!-\! 3}{D \!-\! 2}\Bigr)
\mathbf{P}_{\mu\nu}^{~~\beta\delta} \Bigl[ \square \!-\! 2 H^2\Bigr]
\Bigl[ \square \!-\! D H^2\Bigr] F_{\beta\delta} \; . \qquad
\label{2ndsquare}
\end{eqnarray}
We note in passing that the covariant d`Alembertian commutes with
$\mathbf{P}_{\mu\nu}^{~~\beta\delta}$, just as it did for the
transverse projector $\mathbf{P}_{\mu}^{~\beta}$.

Of course the relevance of the transverse-traceless projector
$\mathbf{P}_{\mu\nu}^{~~~\beta\lambda}$ is that two factors of it
give the exterior operators of the spin two part of the propagator,
\begin{equation}
i\Bigl[ \mbox{}_{\mu\nu} \Delta^2_{\rho\sigma}\Bigr](x;z) = \frac1{4
H^4} \mathbf{P}_{\mu\nu}^{~~\beta\delta}(x) \times
\mathbf{P}_{\rho\sigma}^{~~\lambda\phi}(z) \Bigl[
\mathcal{R}_{\beta\lambda}(x;z) \mathcal{R}_{\delta\phi}(x;z)
\mathcal{S}_2(x;z) \Bigr] \; .
\end{equation}
From the fact that the d'Alembertian commutes with
$\mathbf{P}_{\mu\nu}^{~~\beta\delta}$ we see,
\begin{eqnarray}
\lefteqn{ \frac12 \Bigl[\square \!-\! 2 H^2\Bigr] i\Bigl[
\mbox{}_{\mu\nu} \Delta^2_{\rho\sigma}\Bigr](x;z) } \nonumber \\
& & \hspace{2cm} = \frac1{4 H^4} \mathbf{P}_{\mu \nu}^{~~\beta
\delta}(x) \times \mathbf{P}_{\rho\sigma}^{~~\lambda\phi}(z) \times
\frac12 \Bigl[\square \!-\! 2 H^2\Bigr] \Bigl[ \mathcal{R}_{\beta
\lambda} \mathcal{R}_{\delta \phi} \mathcal{S}_2 \Bigr] \; . \qquad
\end{eqnarray}
The next step is to pass the differential operator through to the
structure function, making use of identities
(\ref{useful1}-\ref{useful2}) from section 4,
\begin{eqnarray}
\lefteqn{\square \Bigl[\mathcal{R}_{\beta\lambda}
\mathcal{R}_{\delta\phi} \mathcal{S}_2\Bigr] =
\mathcal{R}_{\beta\lambda} \mathcal{R}_{\delta\phi} \square
\mathcal{S}_2 + 2 g^{\alpha\gamma}(x) \Bigl[\frac{D
\mathcal{R}_{\beta\lambda}}{D x^{\alpha}} \mathcal{R}_{\delta\phi} +
\mathcal{R}_{\beta\lambda} \frac{D \mathcal{R}_{\delta\phi}}{D
x^{\alpha}} \Bigr] \frac{\partial \mathcal{S}_2}{\partial
x^{\gamma}}} \nonumber \\
& & \hspace{1.5cm} + 2 g^{\alpha\gamma}(x) \frac{D
\mathcal{R}_{\beta\lambda}}{D x^{\alpha}} \frac{D
\mathcal{R}_{\delta \phi}}{D x^{\gamma}} \mathcal{S}_2 + \Bigl[
(\square \mathcal{R}_{\beta\lambda}) \mathcal{R}_{\delta\phi} +
\mathcal{R}_{\beta\lambda} (\square \mathcal{R}_{\delta\phi})\Bigr]
\mathcal{S}_2 \; , \qquad \\
& &  = \mathcal{R}_{\beta\lambda} \mathcal{R}_{\delta\phi}
\Bigl[\square \!+\! 2 H^2\Bigr] \mathcal{S}_2 \nonumber \\
& & \hspace{1cm} + \frac{D}{D x^{\beta}} \Bigl[\frac{\partial
y}{\partial z^{\lambda}} \mathcal{R}_{\delta\phi} \mathcal{S}_2
\Bigr] + \frac{D}{D x^{\delta}} \Bigl[ \mathcal{R}_{\beta\lambda}
\frac{\partial y}{\partial z^{\phi}} \mathcal{S}_2\Bigr] - \frac12
g_{\beta\delta}(x) \frac{\partial y}{\partial z^{\lambda}}
\frac{\partial y}{\partial z^{\phi}} \mathcal{S}_2 \; . \qquad
\label{finalline}
\end{eqnarray}
When the external operators are contracted into this the terms on
the final line of (\ref{finalline}) all drop by virtue of either
transversality or tracelessness. Hence we have,
\begin{equation}
\frac12 \Bigl[\square \!-\! 2 H^2\Bigr] i\Bigl[ \mbox{}_{\mu\nu}
\Delta^2_{\rho\sigma}\Bigr](x;z) = \frac1{4 H^4} \mathbf{P}_{\mu
\nu}^{~~\beta \delta}(x) \times \mathbf{P}_{\rho \sigma}^{~~\lambda
\phi}(z) \Bigl[ \mathcal{R}_{\beta \lambda} \mathcal{R}_{\delta
\phi} \times \frac{\square}2 \mathcal{S}_2 \Bigr] \; .
\label{leftpass}
\end{equation}

It is now time to reduce transverse-traceless projection functional
(\ref{P2}) to standard form (\ref{standard}). Just as we did with
the transverse projection functional of section 4, this is
accomplished by acting $\mathbf{P}_{\alpha\gamma}^{~~\mu\nu}(x)
\times \mathbf{P}_{\kappa\theta}^{~~\rho\sigma}(z)$ on both forms.
When acting on expression (\ref{P2}) tracelessness or transversality
make all but the first term drop out,
\begin{eqnarray}
\lefteqn{\mathbf{P}_{\alpha\gamma}^{~~\mu\nu}(x) \times
\mathbf{P}_{\kappa\theta}^{~~\rho\sigma}(z) \, i\Bigl[
\mbox{}_{\mu\nu} P^2_{\rho\sigma}\Bigr](x;z) } \nonumber \\
& & \hspace{4cm} = \mathbf{P}_{\alpha\gamma}^{~~\mu\nu}(x) \times
\mathbf{P}_{\kappa\theta}^{~~\rho\sigma}(z) \Biggl[ g_{\mu\rho}
g_{\nu\sigma} \frac{i \delta^D(x \!-\! z)}{\sqrt{-g}} \Biggr] \; ,
\qquad \\
& & \hspace{4cm} = \mathbf{P}_{\alpha\gamma}^{~~\mu\nu}(x) \times
\mathbf{P}_{\kappa\theta}^{~~\rho\sigma}(z) \Biggl[
\mathcal{R}_{\mu\rho} \mathcal{R}_{\nu\sigma} \frac{i \delta^D(x
\!-\! z)}{\sqrt{-g}} \Biggr] \; . \qquad \label{rel1}
\end{eqnarray}
On the other hand, acting the same operator on (\ref{standard}), and
making use of relations (\ref{2ndsquare}) and (\ref{leftpass}),
tells us,
\begin{eqnarray}
\lefteqn{\mathbf{P}_{\alpha\gamma}^{~~\mu\nu}(x) \times
\mathbf{P}_{\kappa\theta}^{~~\rho\sigma}(z) \, i\Bigl[
\mbox{}_{\mu\nu} P^2_{\rho\sigma}\Bigr](x;z) = \frac1{4 H^4}
\mathbf{P}_{\alpha\gamma}^{~~ \mu \nu}(x) \times
\mathbf{P}_{\kappa\theta}^{~~\rho\sigma}(z) } \nonumber \\
& & \hspace{3.5cm} \times \Biggl[ \mathcal{R}_{\mu\rho}
\mathcal{R}_{\nu\sigma} \frac14 \Bigl(\frac{D \!-\! 3}{D \!-\!
2}\Bigr)^2 \square^2 \Bigl[ \square \!-\! (D\!-\! 2) H^2\Bigr]^2
\mathcal{P}_2 \Biggr] \; . \qquad \label{rel2}
\end{eqnarray}
Comparing (\ref{rel1}) with (\ref{rel2}) we infer an equation for
the structure function of the transverse-traceless projection
functional,
\begin{equation}
\square^2 \Bigl[ \square \!-\! (D \!-\!2) H^2\Bigr]^2
\mathcal{P}_2(x;z) = 16 H^4 \Bigl( \frac{D \!-\! 2}{D \!-\!
3}\Bigr)^2 \times \frac{i \delta^D(x \!-\! z)}{\sqrt{-g}} \; .
\end{equation}
The solution is easily constructed using relation (\ref{bprop}) and
successive applications of (\ref{Int1}-\ref{Int3}),
\begin{equation}
\mathcal{P}_2(x;z) = \Bigl( \frac{4}{D \!-\! 3} \Bigr)^2 \Biggl[
i\Delta_{AA}(x;z) - 2 i\Delta_{AB}(x;z) + i\Delta_{BB}(x;z) \Biggr]
\; . \label{P2answer}
\end{equation}

The long-sought equation for the spin two structure function derives
from the substitution in equation (\ref{spin2eqn}) of relations
(\ref{leftpass}) and (\ref{P2answer}),
\begin{equation}
\frac12 \square \mathcal{S}_2(x;z) = \Bigl( \frac{4}{D \!-\! 3}
\Bigr)^2 \Biggl[ i\Delta_{AA}(x;z) - 2 i\Delta_{AB}(x;z) +
i\Delta_{BB}(x;z) \Biggr] \; . \label{S2eqn}
\end{equation}
The solution can be found using relations (\ref{Int4}-\ref{Int7b})
from the end of section 3,
\begin{equation}
\mathcal{S}_2(x;z) = \frac{32}{(D \!-\! 3)^2} \Bigl[
i\Delta_{AAA}(x;z) \!-\! 2 i\Delta_{AAB} + i\Delta_{ABB}(x;z) \Bigr]
\; . \label{spin2sol}
\end{equation}

\section{Discussion}

We have constructed the graviton propagator on de Sitter background
in exact de Donder gauge (\ref{deDonder}). Our result takes the form
(\ref{gravdecomp}) of a spin zero part and a spin two part. Both
parts are represented in terms of covariant differential projectors
which automatically enforce the gauge condition, acting on scalar
structure functions. Our form for the spin zero part is given by
relations (\ref{Spin0}) and (\ref{P0op}). The spin two part
(\ref{Spin2}) has a complicated definition involving relations
(\ref{CPdef}-\ref{CDdef3}) and (\ref{Qdef}-\ref{Rdef}). By taking
appropriate traces and commuting differential operators we
eventually derive scalar equations (\ref{S0eqn}) and (\ref{S2eqn})
for the structure functions of the respective parts. These equations
are then solved using the general scalar techniques explained and
summarized in section 3.

We emphasize that our forms for the spin zero and spin two parts of
the propagator involve no assumption about de Sitter invariance, nor
specialization to any particular portion of the de Sitter manifold.
The equations (\ref{S0eqn}) and (\ref{S2eqn}) we derive for the two
structure functions are scalar equations, valid in any coordinate
system and with no inherent assumption about de Sitter invariance.
To emphasize this, we act extra derivatives so as to make the source
on the right hand side proportional to a delta function in each
case,
\begin{eqnarray}
\Bigl[ \square \!+\! D H^2\Bigr] \Bigl[ \square \!+\! 2 (D \!-\! 1)
H^2 \Bigr]^2 \mathcal{S}_0(x;z) & = & -2 \Bigl( \frac{D \!-\! 2}{D
\!-\! 1}\Bigr) \frac{i \delta^D(x \!-\! z)}{\sqrt{-g}} \; , \qquad
\label{EQN0} \\
\square^3 \Bigl[ \square \!-\! (D \!-\! 2) H^2 \Bigr]^2
\mathcal{S}_2(x;z) & = & 32 \Bigl(\frac{D \!-\! 2}{D \!-\! 3}
\Bigr)^2 H^4 \frac{i \delta^D(x \!-\! z)}{\sqrt{-g}} \; . \qquad
\label{EQN2}
\end{eqnarray}
It happens that neither the spin zero structure function
(\ref{spin0sol}) nor its spin two counterpart (\ref{spin2sol}) is de
Sitter invariant. For the spin zero case this is obvious from the
presence of tachyonic mass terms in both of the differential
operators on the left hand side of equation (\ref{EQN0}). The mass
$M_S^2 = -D H^2$ includes a logarithmic singularity which shows up
even in analytic regularization techniques. For the spin two
equation (\ref{EQN2}) the squared operator has positive mass-squared
$M_S^2 = (D-2) H^2$ and would not lead to breaking of de Sitter
invariance were it alone. However, the cubed operator is the same as
that for a massless, minimally coupled scalar --- as might have been
expected from Grishchuk's old result \cite{Grishchuk}. Allen and
Folacci long ago proved that this has no de Sitter invariant
solution \cite{AF}.

Exact de Donder gauge is interesting because de Sitter invariant
constructions based on analytic continuation methods had previously
dismissed it as an infrared divergent special case \cite{Higuchi}.
In fact, all valid gauges show infrared divergences. The special
thing about de Donder gauge is that some of its infrared divergences
are logarithmic so that they are not automatically (and incorrectly)
subtracted by analytic continuation. In all cases the right way to
resolve the infrared divergence is by breaking de Sitter invariance.

We have gone to considerable lengths --- in previous work
\cite{MTW1,MTW2} and again in section 3 --- to elucidate precisely
what goes wrong with previous constructions \cite{INVPROP} which
seemed to produce de Sitter invariant results. However, it worth
pointing out that the fact of de Sitter breaking was already obvious
to cosmologists from the scale invariance of the tensor power
spectrum, which becomes exact in the de Sitter limit \cite{RPW}. It
was also obvious from the explicit form of a propagator constructed
by mode sums on the open submanifold (for which there is no
linearization instability) \cite{TW9,RPW}. On the open submanifold
the $\frac12 D (D+1)$ elements of the de Sitter group break down
into four parts:
\begin{enumerate}
\item{$(D-1)$ spatial translations;}
\item{$\frac12 (D-2)(D-1)$ spatial rotations;}
\item{A single dilatation; and}
\item{$(D-1)$ spatial special conformal transformations.}
\end{enumerate}
The gauge condition only breaks the last of these, but the solution
for the propagator additionally breaks dilatation invariance
\cite{TW9,RPW}. The physical de Sitter breaking of this propagator
was demonstrated by Kleppe, who augmented a naive de Sitter
transformation by the compensating gauge transformation needed to
restore the gauge condition \cite{Kleppe}. Had the propagator been
physically invariant this technique would have revealed it.

We should also comment on the apparent conflict of our result with
the pro-invariance argument given by Marolf and Morrison
\cite{DMIAM}, based on work by Higuchi \cite{Higuchi2}. They dealt
with free dynamical gravitons in a noncovariant gauge on the full de
Sitter manifold and they were able to construct the complete panoply
of mode solutions and inner products. This should imply a vacuum
which is physically de Sitter invariant --- that is, invariant once
the compensating gauge transformation is included. We know of no
problem with this work but it should be noted that the propagator
one gets using only dynamical gravitons (that is, the spatial,
transverse-traceless polarizations) is not complete. It is like the
purely spatial and transverse photon propagator of flat space
electrodynamics in Coulomb gauge. To fully describe electromagnetic
interactions also requires the instantaneous Coulomb interaction.
Both of these are part of the same propagator in a covariant gauge
such as the one we employ here.

The constrained, spin zero part of our propagator --- which is
missing from the transverse-traceless part --- provides the largest
source of the de Sitter breaking we found. It is relatively simple
to show that the de Sitter breaking terms in $\mathcal{S}_0(x;z)$ do
not drop out when acted upon by the spin zero projector
$\mathcal{P}_{\mu\nu}(x) \times \mathcal{P}_{\rho \sigma}(z)$. The
spin two structure function contains less severe de Sitter breaking
terms of the form,
\begin{equation}
\Bigl[\mathcal{S}_2(x;z) \Bigr]_{\rm de\ Sitter\ \atop breaking} =
\sum_{k=1}^3 s_k \Bigl[ \ln(a_x a_z) \Bigr]^k \; .
\end{equation}
It is possible that these drop out from the spin two part of the
propagator (\ref{Spin2}) after all eight of the derivatives have
been taken. In that case our work would be fully consistent with
that of Marolf and Morrison. However, what we expect is that one of
the infrared logarithms survives, which seems to be indicated by the
scale invariance of the tensor power spectrum.

The fact of de Sitter breaking in this system cannot be disputed,
but there is wide freedom as to how one chooses to manifest that
breaking. This freedom amounts to picking the initial state. We have
chosen the explicit solutions of section 3 so as to preserve the
symmetries of homogeneity and isotropy, which allow one to view de
Sitter as a special case of a spatially flat, FRW geometry. This
choice is known in the literature as the ``$E(3)$ vacuum.'' Readers
who prefer to preserve another subgroup can do so by starting from
our scalar equations (\ref{EQN0}) and (\ref{EQN2}).

We wrote this paper to help resolve the long-standing controversy
about de Sitter breaking for free gravitons, however, it has other
applications. One of these is to test for gauge dependence in
quantum gravitational loop corrections from primordial inflation. Of
course gauge-fixed Green's functions will show such dependence,
mingled with valid physical information. In flat space we would sift
out the gauge dependence by forming the S-matrix. That observable is
not available in cosmology \cite{Witten}, and there is not yet any
consensus for what replaces it. One technique is simply to carry out
computations in different gauges. It may be that the leading
infrared logarithm contributions (e.g., the one loop contribution to
the fermion field strength from inflationary gravitons \cite{MW2})
are independent of the choice of gauge. Now we can test this
conjecture using a completely different gauge from the one
\cite{TW9,RPW} employed in all previous computations.

Our propagator should also make renormalization simpler because it
precludes the appearance of noninvariant counterterms. These
complicated the analysis for previous computations \cite{MW2,KW2}.
It may also be that the gauge condition (\ref{deDonder}) and the
special properties of the differential projectors in our propagator
make actual computations simpler. That turned out to be the case
with the vector propagator in Lorentz gauge \cite{TW2} for a variety
of one and two loop computations \cite{PTW1,PTW2,PTW3}.

\centerline{\bf Acknowledgements}

This work was partially supported by FQXi Mini Grant \#MGB-08-008,
by NWO Veni Project \# 680-47-406, by European Union Grant
FP-7-REGPOT-2008-1-CreteHEPCosmo-228644, by NSF grant PHY-0855021,
and by the Institute for Fundamental Theory at the University of
Florida.


\begin{thebibliography}{99}

\bibitem{MTW1} S. P. Miao, N. C. Tsamis and R. P. Woodard, J. Math.
Phys. {\bf 50} (2009) 122502, arXiv:0907.4930.

\bibitem{INVPROP} B. Allen and M. Turyn, Nucl. Phys. {\bf B292} (1987)
813; S. W. Hawking, T. Hertog and N. Turok, Phys. Rev. {\bf D62}
(2000) 063502, hep-th/0003016; A. Higuchi and S. S. Kouris, Class.
Quant. Grav. {\bf 18} (2001) 4317, gr-qc/0107036; A. Higuchi and R.
H. Weeks, Class. Quant. Grav. {\bf 20} (2003) 3006, gr-qc/0212031.

\bibitem{MTW2} S. P. Miao, N. C. Tsamis and R. P. Woodard, J. Math.
Phys. {\bf 51} 072503, arXiv:1002.4037.

\bibitem{TW1} N. C. Tsamis and R. P. Woodard, Class. Quant. Grav.
{\bf 18} (2001) 83, hep-ph/0007166.

\bibitem{Marolf} D. Marolf, private communication on July 6, 2010.

\bibitem{AJ} B. Allen and T. Jacobson, Commun. Math. Phys. {\bf 103}
(1986) 669.

\bibitem{KW1} E. O. Kahya and R. P. Woodard, Phys. Rev. {\bf D72}
(2005) 104001, gr-qc/0508015; Phys. Rev. {\bf D74} (2006) 084012,
gr-qc/0608049.

\bibitem{RPW} R. P. Woodard, ``de Sitter breaking in field theory,''
in {\it Deserfest: A celebration of the life and works of Stanley
Deser} (World Scientific, Hackensack, 2006) eds. J. T. Liu, M. J.
Duff, K. S. Stelle and R. P. Woodard, p. 339, gr-qc/0408002.

\bibitem{TW2} N. C. Tsamis and R. P. Woodard, J. Math. Phys. {\bf
48} (2007) 052306, gr-qc/0608069.

\bibitem{PTW1} T. Prokopec, N. C. Tsamis and R. P. Woodard, Class.
Quant. Grav. {\bf 24} (2007) 201, gr-qc/0607094.

\bibitem{IAEM} I. Antoniadis and E. Mottola, J. Math. Phys. {\bf
32} (1991) 1037.

\bibitem{Folacci} A. Folacci, Phys. Rev. {\bf D46} (1992) 2553,
arXiv:0911.2064; Phys. Rev. {\bf D53} (1996) 3108.

\bibitem{Higuchi} A. Higuchi and Y. C. Lee, Phys. Rev. {\bf D78} (2008)
084031, \hfil\break arXiv:0808.0642; M. Faizal and A. Higuchi, Phys.
Rev. {\bf D78} (2008) 067502, arXiv:0806.3735.

\bibitem{WMAP} E. Komatsu, Astrophys. J. Suppl. {\bf 192} (2011) 18,
arXiv:1001.4538.

\bibitem{KOW} E. O. Kahya, V. K. Onmeli and R. P. Woodard, Phys.
Lett. {\bf B694} (2010) 101, arXiv:1006.3999.

\bibitem{old} E. Schr\"odinger, Physica {\bf 6} (1939) 899; T.
Imamura, Phys. Rev. {\bf 118} (1960) 1430; L. Parker, Phys. Rev.
Lett. {\bf 21} (1968) 562; Phys. Rev. {\bf 183} (1969) 1057; Phys.
Rev. {\bf D3} (1971) 346.

\bibitem{Parker} L. H. Ford and L. Parker, Phys. Rev. {\bf D16}
(1977) 1601.

\bibitem{Grishchuk} L. P. Grishchuk, Sov. Phys. JETP {\bf 40} (1975)
409.

\bibitem{Starobinsky} A. A. Starobinsky, JETP Lett. {\bf 30} (1979)
682.

\bibitem{Mukhanov} V. F. Mukhanov and G. V. Chibisov, JETP Lett.
{\bf 33} (1981) 532.

\bibitem{OW} V. K. Onemli and R. P. Woodard, Class. Quant. Grav.
{\bf 19} (2002) 4607, qc-gr/0204065; Phys. Rev. {\bf D70} (2004)
107301, gr-qc/0406098.

\bibitem{BOW} T. Brunier, V. K. Onemli and R. P. Woodard, Class. Quant.
Grav. {\bf 22} (2005) 59, gr-qc/0408080; E. O. Kahya and V. K.
Onemli, Phys. Rev. {\bf D76} (2007) 043512, gr-qc/0612026.

\bibitem{PWGP} T. Prokopec and R. P. Woodard, JHEP {\bf 0310} (2003)
059, astro-ph/0309593; B. Garbrecht and T. Prokopec, Phys. Rev. {\bf
D73} (2006) 064036, gr-qc/0602011.

\bibitem{DW} L. D. Duffy and R. P. Woodard, Phys. Rev. {\bf D72} (2005)
024023, hep-ph/0505156.

\bibitem{MW1} S. P. Miao and R. P. Woodard, Phys. Rev. {\bf D74} (2006)
044019, gr-qc/0602110.

\bibitem{PTW} T. Prokopec, O. Tornkvist and R. P. Woodard, Phys. Rev.
Lett. {\bf 89} (2002) 101301, astro-ph/0205331; Annals Phys. {\bf
303} (2003) 251, gr-qc/0205130; T. Prokopec and R. P. Woodard,
Annals Phys. {\bf 312} (2004) 1, gr-qc/0310056.

\bibitem{PTW2} T. Prokopec, N. C. Tsamis and R. P. Woodard,
Phys. Rev. {\bf D78} (2008) 043523, arXiv:0802.3673.

\bibitem{PTW3} T. Prokopec, N. C. Tsamis and R. P. Woodard, Annals Phys.
{\bf 323} (2008) 1324, arXiv:0707.0847.

\bibitem{TW3} N. C. Tsamis and R. P. Woodard, Ann. Phys. {\bf 321}
(2006) 875, gr-qc/0506089.

\bibitem{MW2} S. P. Miao and R. P. Woodard, Class. Quant. Grav. {\bf
23} (2006) 1721, gr-qc/0511140; Phys. Rev. {\bf D74} (2006) 024021,
gr-qc/0603135; S. P. Miao, arXiv:0705.0767.

\bibitem{MW3} S. P. Miao and R. P. Woodard, Class. Quant. Grav. {\bf
25} (2008) 145009, arXiv:0803.2377.

\bibitem{KW2} E. O. Kahya and R. P. Woodard, Phys. Rev. {\bf D76}
(2007) 124005, arXiv:0709.0536; Phys. Rev. {\bf D77} (2008) 084012,
arXiv:0710.5281.

\bibitem{PW} S. Park and R. P. Woodard, Phys. Rev. {\bf D83} (2011)
084049, arXiv:1101.5804.

\bibitem{TW4} N. C. Tsamis and R. P. Woodard, Nucl. Phys. {\bf B724}
(2005) 295, gr-qc/0505115.

\bibitem{KK} H. Kitamoto and Y. Kitazawa, arXiv:1012.5930.

\bibitem{Ford} L. H. Ford, Phys. Rev. {\bf D31} (1985) 710.

\bibitem{Finelli} F. Finelli, G. Marozzi, G. P. Vacca and G.
Venturi, Phys. Rev. {\bf D71} (2005) 023522, gr-qc/0407101.

\bibitem{TW5} N. C. Tsamis and R. P. Woodard, Nucl. Phys. {\bf B474}
(1996) 235, hep-ph/960215; Ann. Phys. {\bf 253} (1997) 1,
hep-ph/9602316.

\bibitem{TW6} N. C. Tsamis and R. P. Woodard, Phys. Rev. {\bf D54}
(1996) 2621, hep-ph/960217.

\bibitem{Albert} G. Perez-Nadal, A. Roura and E. Verdaguer, JCAP
{\bf 1005} (2010) 036, arXiv:0911.4870.

\bibitem{back} V. F. Mukhanov, L. R. W. Abramo and R. H.
Brandenberger, Phys. Rev. Lett. {\bf 78} (1997) 1624, gr-qc/9609026;
L. R. W. Abramo, R. H. Brandenberger and V. F. Mukhanov, Phys. Rev.
{\bf D56} (1997) 3248, gr-qc/9704037; L. R. W. Abramo and R. P.
Woodard, Phys. Rev. {\bf D60} (1999) 044010, astro-ph/9811430; Phys.
Rev. {\bf D60} (1999) 044011, astro-ph/9811431; A. Ghosh, R. Madden
and G. Veneziano, Nucl. Phys. {\bf B570} (2000) 207, hep-th/9908024.

\bibitem{measure} W. Unruh, astro-ph/9802323; L. R. Abramo and R. P.
Woodard, Phys. Rev. {\bf D65} (2002) 043507, astro-ph/0109271; Phys.
Rev. {\bf D65} (2002) 063515, astro-ph/0109272; G. Geshnizjani and
R. Brandenberger, Phys. Rev. {\bf D66} (2002) 123507, gr-qc/0204074;
JCAP {\bf 0504} (2005) 006, hep-th/0310265.

\bibitem{puremeasure} N. C. Tsamis and R. P. Woodard, Class. Quant.
Grav. {\bf 22} (2005) 4171, gr-qc/0506056; Phys. Rev. {\bf D78}
(2008) 028501, arXiv:0708.2004; Class. Quant. Grav. {\bf 26} (2009)
105006, arXiv:0807.5006; J. Garriga and T. Tanaka, Phys. Rev. {\bf
D77} (2008) 024021, arXiv:0706.0295.

\bibitem{many} J. Maldacena, JHEP {\bf 0305} (2003) 013, astro-ph/0210603;
D. Boyanovsky, H. J. de Vega and N. G. Sanchez, Nucl. Phys. {\bf
B747} (2006) 25, astro-ph/0503669; Phys. Rev. {\bf D72} (2005)
103006, astro-ph/0507596; M. Sloth, Nucl. Phys. {\bf B748} (2006)
149, astro-ph/0604488; Nucl. Phys. {\bf B775} (2007) 78,
hep-th/0612138; A. Biland\v{z}i\'c and T. Prokopec, Phys. Rev. {\bf
D76} (2007) 103507, arXiv:0704.1905; M. van der Meulen and J. Smit,
JCAP {\bf 0711} (2007) 023, arXiv:0707.0842; D. H. Lyth, JCAP {\bf
0712} (2007) 016, arXiv:0707.0361; D. Seery, JCAP {\bf 0711} (2007)
025, arXiv:0707.3377; JCAP {\bf 0802} (2008) 006, arXiv:0707.3378;
JCAP {\bf 0905} (2009) 021, arXiv:0903.2788; N. Bartolo, S.
Matarrese, M. Pietroni, A. Riotto and D. Seery, JCAP {\bf 0801}
(2008) 015, arXiv:0711.4263, Y. Urakawa and K. I Maeda, Phys. Rev.
{\bf D78} (2008) 064004, arXiv:0801.0126, A. Riotto and M. Sloth,
JCAP {\bf 0804} (2008) 030, arXiv:0801.1845, K. Enqvist, S. Nurmi,
D. Podolsky and G. I. Rigopoulos, JCAP {\bf 0804} (2008) 025,
arXiv:0802.0395, P. Adshead, R. Easther and E. A. Lim, Phys. Rev.
{\bf D79} (2009) 063504, arXiv:0809.4008, J. Kumar, L. Leblond and
A. Rajaraman, JCAP {\bf 1004} (2010) 024, arXiv:0909.2040; C. P.
Burgess, R. Holman, L. Leblond and S. Shandera, JCAP {\bf 1003}
(2010) 033, arXiv:0912.1608, L. Senatore and M. Zaldarriaga, JHEP
{\bf 1012} (2010) 008, arXiv:0912.2734; D. Seery, Class. Quant.
Grav. {\bf 27} (2010) 124005, arXiv:1005.1649; S. B. Giddings and M.
S. Sloth, JCAP {\bf 1101} (2011) 023, arXiv:1005.1056; JCAP {\bf
1007} (2010) 015, arXiv:1005.3287, arXiv:1104.0002.

\bibitem{gauge} Y. Urakawa and T. Tanaka, Prog. Theor. Phys. {\bf
122} (2009) 779, arXiv:0902.3209, Prog. Theor. Phys. {\bf 122}
(2010) 1207, arXiv:0904.4415; Phys. Rev. {\bf D82} (2010) 121301,
arXiv:1007.0468; JCAP {\bf 1105} (2011) 014, arXiv:1103.1251,
arXiv:1103.1251; C. T. Byrnes, M. Gerstenlauer, A. Hebecker, S.
Nurmi and G. Tasinato, JCAP {\bf 1008} (2010) 006, arXiv:1005.3307,
M. Gerstenlauer, A. Hebecker and G. Tasinato, arXiv:1102.0560; Y.
Urakawa, arXiv:1105.1078.

\bibitem{Weinberg} S. Weinberg, Phys. Rev. {\bf D72} (2005) 043514,
hep-th/0506236; Phys. Rev. {\bf D74} (2006) 023508, hep-th/0605244;
K. Chaicherdsakul, Phys. Rev. {\bf D75} (2007) 063522,
hep-th/0611352.

\bibitem{UVIR} S. Weinberg, Phys. Rev. {\bf D83} (2011) 063508,
arXiv:1011.1630; W Xue, K. Dasgupta and R. Brandenberger, Phys. Rev.
{\bf D83} (2011) 083520, arXiv:1103.0285.

\bibitem{Polyakov} A. M. Polyakov, Sov. Phys. Usp. {\bf 25} (1982) 187;
Nucl. Phys. {\bf B834} (2010) 316, arXiv:0912.5503; D. Krotov and A.
M. Polyakov, Nucl. Phys. {\bf B849} (2011) 410, arXiv:1012.2107.

\bibitem{oldclaims} A. M. Polyakov, Sov. Phys. Usp. {\bf 25} (1982)
187; N. P. Myhrvold, Phys. Rev. {\bf D28} (1983) 2439; E. Mottola,
Phys. Rev. {\bf D31} (1985) 754; Phys. Rev. {\bf 33} (1986) 2136; P.
O. Mazur and E. Mottola, Nucl. Phys. {\bf B278} (1986) 694; I.
Antoniadis, J. Iliopoulos, and T. N. Tomaras, Phys. Rev. Lett. {\bf
56} (1986) 1319; N. C. Tsamis and R. P. Woodard, Phys. Lett. {\bf
B301} (1993) 351; A. D. Dolgov, M. B. Einhorn and V. I Zakharov,
Phys. Rev. {\bf D52} (1995) 717, gr-qc/9403056.

\bibitem{destab} A. Higuchi, Class. Quant. Grav. {\bf 26} (2009) 072001,
arXiv:0809.1255; E. T. Akhmedov, Mod. Phys. Lett. {\bf A25} (2010)
2815, arXiv:0909.3722; E. T. Akhmedov and P. Burda, Phys. Lett. {\bf
B687} (2010) 267, arXiv:0912.3435; E. T. Akhmedov, A. Roura and A.
Sadofyev, Phys. Rev. {\bf D82} (2010) 044035, arXiv:1006.3274; D.
Marolf, I. A. Morrison, Phys. Rev. {\bf D82} (2010) 105032,
arXiv:1006.0035; arXiv:1010.5327; arXiv:1104.4343; H. Kitamoto and
Y. Kitazawa, Nucl. Phys. {\bf B839} (2010) 552, arXiv:1004.2451; A.
Higuchi, D. Marolf and I. A. Morrison, Phys. Rev. {\bf D83} (2011)
084029, arXiv:1012.3415; C. P. Burgess, R. Holman, L. Lelond, S.
Shandera, JCAP {\bf 1010} (2010) 017, arXiv:1005.3551; D.
Boyanovsky, R. Holman, JHEP {\bf 1105} (2011) 047.

\bibitem{BD} T. S. Bunch and P. C. W. Davies, Proc. R. Soc. {\bf A357}
(1977) 381.

\bibitem{CT} N. A. Chernikov and E. A. Tagirov, Annales Poincare Phys. Theor.
{\bf A9} (1968) 109.

\bibitem{JMPW1} T. M. Janssen, S. P. Miao, T. Prokopec and R. P. Woodard,
Class. Quant. Grav. {\bf 25} (2008) 245013, arXiv:0808.2449.

\bibitem{FP} L. H. Ford and L. Parker, Phys. Rev. {\bf D16} (1977) 245.

\bibitem{classic} A. Vilenkin and L. H. Ford, Phys. Rev. {\bf D26} (1982) 1231;
A. D. Linde, Phys. Lett. {\bf 116B} (1982) 335; A. A. Starobinsky,
Phys. Lett. {\bf 117B} (1982) 175.

\bibitem{AF} B. Allen and A. Folacci, Phys. Rev. {\bf 35} (1987)
3771.

\bibitem{AV} A. Vilenkin, Nucl. Phys. {\bf B226} (1983) 527.

\bibitem{TW7} N. C. Tsamis and R. P. Woodard, Class. Quant. Grav. {\bf 11}
(1994) 2969.

\bibitem{JMPW2} T. M. Janssen, S. P. Miao, T. Prokopec and R. P. Woodard,
JCAP {\bf 0905} (2009) 003, arXiv:0904.1151.

\bibitem{TW8} N. C. Tsamis and R. P. Woodard, Phys. Lett. {\bf B292}
(1992) 269.

\bibitem{TW9} N. C. Tsamis and R. P. Woodard, Commun. Math. Phys. {\bf 162}
(1994) 217.

\bibitem{Kleppe} G. Kleppe, Phys. Lett. {\bf B317} (1993) 305.

\bibitem{DMIAM} D. Marolf and I. A. Morrison, Class. Quant. Grav.
{\bf 26} (2009) 235003, arXiv:0810.5163.

\bibitem{Higuchi2} A. Higuchi, Class. Quant. Grav. {\bf 8} (1991)
1983; Class. Quantu. Grav. {\bf 8} (1991) 2005.

\bibitem{Witten} E. Witten, hep-th/0106109; A Strominger, JHEP
{\bf 0110} (2001) 034, hep-th/0106113.


\end{thebibliography}
\end{document}